\documentclass[aps,prd,showpacs,preprintnumbers,nofootinbib,twocolumn]{revtex4}
\usepackage{bm}
\usepackage{latexsym}
\usepackage{dcolumn}
\usepackage{amsfonts,amssymb}
\usepackage{graphicx,epsfig}
\usepackage{psfrag}
\usepackage{amsmath,amssymb}

\def\be{\begin{equation}}
\def\ee{\end{equation}}

\def\a  {\alpha}
\def\b  {\beta}

\def\d  {\delta}
\def\ep {\epsilon}

\def\k  {\kappa}

\def\m  {\mu}
\def\n  {\nu}
\def\o  {\omega}

\def\s {\sigma}

\def\ve{\varepsilon}
\def\vk{\varkappa}
\def\vp{\varphi}

\def\vr{\varrho}
\def\vs{\varsigma}

\def\beq{\begin{equation}}
\def\eeq{\end{equation}}
\def\br{\begin{eqnarray}}
\def\er{\end{eqnarray}}
\def\benu{\begin{enumerate}}
\def\eenu{\end{enumerate}}
\def\nn{\nonumber} 
\def\pa{{\partial}}
\def\l{\left}
\def\r{\right}    
\def\ov{\overline}
\def\wt{\widetilde}
\def\Hbar{\mathcal H}

\def\MPl{M_{_{\rm Pl}}^2}
\def\MPlh{M_{_{\rm Pl}}}

\def\DD{{\cal D}}
\begin{document}

\preprint{hep-th/0507086}

\title{Gauge-invariant perturbation theory for trans-Planckian inflation}
\author{S.~Shankaranarayanan}
\email[]{E-mail: shanki@ictp.trieste.it}
\author{Musongela  Lubo}
\email[]{E-mail: muso@ictp.trieste.it}
\affiliation{HEP Group, International Centre for Theoretical Physics,\\
Strada costiera 11, 34100 Trieste, Italy.}

\date{\today}

%%%%%%%%%%%%%%%%%%%%%%%%%%%%%%%%%%%%%%%%%%%%%%%%%%%%%%%%%%%%%%%%%%%%%%%%%%
%%%%%%%%%%%%%%%%%%%%%%%%%% ABSTRACT %%%%%%%%%%%%%%%%%%%%%%%%%%%%%%%%%%%%%%
%%%%%%%%%%%%%%%%%%%%%%%%%%%%%%%%%%%%%%%%%%%%%%%%%%%%%%%%%%%%%%%%%%%%%%%%%%

\begin{abstract}

The possibility that the scale-invariant inflationary spectrum may be
modified due to the hidden assumptions about the Planck scale physics
--- dubbed as trans-Planckian inflation --- has received considerable
attention. To mimic the possible trans-Planckian effects, among
various models, modified dispersion relations have been popular in the
literature. In almost all the earlier analyzes, unlike the canonical
scalar field driven inflation, the trans-Planckian effects are
introduced to the scalar/tensor perturbation equations in an {\it ad
hoc manner} -- without calculating the stress-tensor of the
cosmological perturbations from the covariant Lagrangian.  In this
work, we perform the gauge-invariant cosmological perturbations for
the single-scalar field inflation with the Jacobson-Corley dispersion
relation by computing the fluctuations of all the fields including the
unit time-like vector field which defines a preferred rest frame. We
show that: (i) The non-linear effects introduce corrections {\it only}
to the perturbed energy density. The corrections to the energy density
vanish in the super-Hubble scales. (ii) The scalar perturbations, in
general, are not purely adiabatic. (iii) The equation of motion of the
Mukhanov-Sasaki variable corresponding to the inflaton field is
different than those presumed in the earlier analyzes. (iv) The tensor
perturbation equation remains unchanged. We perform the classical
analysis for the resultant system of equations and also compute the
power-spectrum of the scalar perturbations in a particular limit. We
discuss the implications of our results and compare with the earlier
results.
\end{abstract}
\maketitle

\section{Introduction} 

Cosmological inflation is currently considered to be the best paradigm
for describing the early stages of the universe
\cite{Linde:1990-bk,Liddle-Lyth:2000-bk}. 
Inflation leads to the existence of a causal mechanism for producing
fluctuations on cosmological scales (when measured today), which at
the time of matter-radiation equality had a physical wavelength larger
than the Hubble radius. Thus, it solves several conceptual problems of
standard cosmology and leads to a predictive theory of the origin of
cosmological fluctuations.

During inflation, the physical wavelength corresponding to a fixed
comoving scale decreases exponentially as time decreases whereas the
Hubble radius is constant. Thus, as long as the period of inflation is
sufficiently long, all scales of cosmological interest today originate
inside the Hubble radius during inflation. Recently, it has been
realized that if inflation lasts slightly longer than the minimal time
(i. e. the time it needs to last in order to solve the horizon problem
and to provide a causal generation mechanism for CMB fluctuations),
then the corresponding physical wavelength of these fluctuations at
the beginning of inflation will be smaller than the Planck
length. This is commonly referred to as the {\it trans-Planckian
problem of inflation}
\cite{Brandenberger-Mart:2000,Martin-Bran:2000,Niemeyer:2000}. 

%Due to this possibility
Naturally, considerable amount of attention has been devoted to
examine the possibility of detecting trans-Planckian imprints on the
CMB
\cite{Tanaka:2000,Niemeyer-Pare:2001,Kempf-Niem:2001,Starobinsky:2001,Easther-Green:2001a,Lemoine-Lubo:2001,Bastero-Gil-Fram:2001,Easther-Green:2001b,Shanki:2002,Danielsson:2002a,Brandenberger-Ho:2002,Hassan-Sloth:2002,Easther-Green:2002,Kaloper-Kleb:2002,Martin-Bran:2003,Bastero-Gil-Free:2003,Shanki-Srir:2004a,Shanki-Srir:2004b,Brandenberger-Mart:2004,Tsujikawa-Maar:2003,Sriram-paddy:2004,Greene-Scha:2004,Calcagni:2004a}. 
Broadly, there have been two approaches in the literature in order to
study these effects. In the first approach, the specific nature of
trans-Planckian physics is not presumed, but is rather described by the
boundary conditions imposed on the mode at the cut-off scale. In the
second approach, which is of interest in this work, one incorporates
quantum gravitational effects by introducing the fundamental length
scale into the standard field theory in a particular fashion.

In almost all the analyses performed so-far in the literature, the
trans-Planckian effects are introduced into the scalar and tensor
perturbation equations in an {\it ad hoc manner}. In the standard
inflation, the scalar and tensor perturbation equations are derived,
from the first principles, in a gauge-invariant manner
\cite{Kodama-Sasa:1984,Mukhanov-Feld:1992}. However in the
trans-Planckian inflationary scenario, to our knowledge, such a
calculation has never been performed and the scalar/tensor 
power spectrum, with the trans-Planckian corrections, were 
obtained from the presumed perturbation equations. 

With the possibility of detecting the trans-Planckian signatures in
the current and future CMB experiments\footnote{For the current status
and developments of WMAP and PLANCK, see the following URLs:
http://map.gsfc.nasa.gov/,
http://astro.estec.esa.nl/SA-general/Projects/Planck}, it becomes {\it
imperative} to obtain the scalar and tensor perturbation equation from
the first principles. (For recent work on the trans-Planckian
constraints from the CMB, see
Refs. \cite{Martin-Ring:2003a,Martin-Ring:2004a,Martin-Ring:2004b,Easther-Kinn:2004,Easther-Kinn:2005}.)  In this work, we
perform the gauge-invariant cosmological perturbation theory for the
single-scalar field inflation with the trans-Planckian
corrections. The model we shall consider in this work is a
self-interacting scalar field in (3 + 1)-dimensional space-time
satisfying a linear wave equation with higher spatial derivative
terms. The dispersion relation [$\omega = \omega(k)$] thus differs at
high wave-vectors from that of the ordinary wave equation. Such a
model breaks the local Lorentz invariance explicitly while preserving
the rotational and translational invariance.  The particular
dispersion relation we shall study in detail is
\beq
\o^2(k)=|\vec{k}|^2 + b_{11} |\vec{k}|^4,
\label{eq:dr}
\eeq
where $b_{11}$ is a dimensionfull parameter. $b_{11} < 0 $ implies
subluminal group velocity, while $b_{11} >0$ implies superluminal
group velocity. The above dispersion relation is a subset of a general
class of the form $\o^2=|\vec{k}|^2[1+g(|\vec{k}|/k_0)]$, where $g$ is
a function which vanishes as $k_0 \to \infty$, and $k_0$ is a constant
which sets the scale for the deviation from Lorentz invariance.  It
has been suggested that these general modified dispersion relation
might arise in loop quantum gravity
\cite{Gambini-Pull:1998,Alfaro-Mora:1999}, or more generally from an
unspecified modification of the short distance structure of space-time
(see for example Refs. \cite{Amelino-Camelia:1999,Jacobson:1999}).
Possible observational consequences have also been studied. For an
up-to-date review, see Refs. \cite{Mattingly:2005,Jacobson-Libe:2005}.

Even though the above dispersion relation breaks the local Lorentz
invariance explicitly it has been shown that it is possible to write
a Lagrangian for the above field in a generally covariant manner,
consistent with spatial translation and rotation invariance, by
introducing a unit time-like Killing vector field which defines a
particular direction
\cite{Jacobson-Matt:2000,Lemoine-Lubo:2001,Lim:2004} (see
Sec. (\ref{sec:MDR}) for more details). In this work, we use such a
framework in-order to obtain the scalar/tensor perturbation equations
for such a model during inflation.

Using the covariant Lagrangian used in Ref. \cite{Jacobson-Matt:2000},
we obtain the perturbed stress-tensor for the scalar and tensor
perturbations about the FRW background. We show that: (i) The
non-linear effects introduce corrections to the perturbed energy
density while the other components of the stress-tensor remains
unchanged. (ii) The non-linear terms contributing to the stress-tensor
are proportional to $k^2$. Hence in the super-Hubble scales the
trans-Planckian contributions to the perturbed energy density, as
expected, can be ignored. (iii) The spatial higher derivative terms
appear {\it only} in the equation of the motion of the perturbed
inflation field ($\d\vp$) and {\it not} in the equation of motion of
the scalar perturbations ($\Phi$). (iv) Unlike the canonical scalar
field inflation, the perturbations, in general, are {\it not} purely
adiabatic. The entropic perturbations generated during the inflation,
however, vanish at the super-Hubble scales. The speed of propagation
of the perturbations is a constant and is less than the speed of
light.  (v) The tensor perturbation equation remain unchanged
indicating that the well-know consistency relation between the scalar
and tensor ratio will also be broken in this model.

We obtain the equation of motion of the Mukhanov-Sasaki variable
corresponding to the inflaton field. We show that the equation of
motion derived from the gauge-invariant perturbation theory is {\it not}
same as those {\it assumed} in the earlier analyzes
\cite{Brandenberger-Mart:2000,Martin-Bran:2000,Niemeyer:2000}. 
Later, we combine the system of differential equations into a single
differential equation in $\Phi$ and obtain the solutions for the
power-law inflation in different regimes. We also obtain the spectrum
of scalar perturbations in a particular limit and compare with the
earlier results.

This paper is organized as follows: In the following section, the
theory of cosmological perturbations for the canonical single scalar
field inflation is discussed and essential steps leading to the
perturbation equation are reviewed. In Sec. (\ref{sec:MDR}), we
discuss the general covariant formulation of the Lagrangian describing
the scalar field with modified dispersion relation and derive the
corresponding stress-tensor. In Sec. (\ref{sec:per-st}), we obtain the
perturbed stress-tensor for the scalar and tensor perturbations.  In
Sec. (\ref{sec:Sca-Per-Eq}), we obtain the scalar perturbation
equation and the equation of motion of the Mukhanov-Sasaki
variable. In Sec. (\ref{sec:clas-ana}), we perform the classical
analysis and obtain the form of the scalar perturbations ($\Phi$) in
the various regimes. In Sec. (\ref{sec:Pow-Spe}), we solve the
perturbation equations in a particular limit and obtain the
power-spectrum of the perturbations. Our results are summarized and
discussed in the last section. In Appendices (A, B) we derive the
equations of motion of the fields in the FRW and perturbed FRW
backgrounds. In Appendix (C), we obtain the equation of motion of the
Bardeen potential in our model.

Through out this paper, the metric signature we adopt is $(+,-,-,-)$
\cite{Landau-Lifs:1976-bk2}, we set $\hbar~ =~ c = ~1$ and $1/(8
\pi G) = \MPl$. The various physical quantities with the {\it
over-line} refers to the values evaluated for the homogeneous and
isotropic FRW background. A dot denotes derivative with respect to the
cosmic time ($t$), a prime stands for a derivative with respect to
conformal time ($\eta$) and $_{,i}$ denotes a derivative w.r.t space
components.  We follow the notation of Ref. \cite{Mukhanov-Feld:1992}
to provide easy comparison.

%%%%%%%%%%%%%%%%%%%%%%%%%%%%%%%%%%%%%%%%%%%%%%%%%%%%%%%%%%%%%%%%%%%%%%%%%%
%%%%%%%%%%%%%%%%%%%%%%%%%  NEW SECTION %%%%%%%%%%%%%%%%%%%%%%%%%%%%%%%%%%%
%%%%%%%%%%%%%%%%%%%%%%%%%%%%%%%%%%%%%%%%%%%%%%%%%%%%%%%%%%%%%%%%%%%%%%%%%%

\section{Gauge-invariant perturbation: Canonical single field inflation}
\label{sec:GIV-Per}

In this section, we obtain the scalar and tensor perturbation
equations for the canonical single scalar field inflation.  In the
following subsection, we discuss key properties of the perturbed FRW
metric and the ``gauge problem'' of the scalar perturbations. In the
subsequent subsections, we discuss the matter Lagrangian and provide
key steps in obtaining the scalar/tensor perturbation equations.

\subsection{Perturbed FRW metric}

We consider perturbations about a spatially flat $(3 + 1)$-dimensional
FRW line element
\br
ds^2 &= &\ov{g}_{\m\nu} dx^{\mu} dx^{\nu} \nn \\
&=& dt^2-a^{2}(t)\, d{\bf x}^2 = 
a^{2}(\eta)\l(d\eta^{2} - d{\bf x}^2\r),
\label{eq:frw}
\er
where $t$ is the cosmic time, $a(t)$ is the scale factor and
$\eta=\int \l[dt/a(t)\r]$ denotes the conformal time.  $H$ is the
Hubble parameter given by $H \equiv \dot{a}/{a}$ while $\Hbar \equiv
a'/a$ is related to the Hubble parameter by the relation $\Hbar =
H\,a$.

At the linear level, for the canonical single scalar field inflation,
the metric perturbations ($\delta g_{\mu\nu}$) can be categorized into
two distinct types --- scalar and tensor perturbations. Thus, the
perturbed FRW line-element can be written as
\br
\label{eq:per-frw}
{\rm d}s^2 &=& a^2(\eta )\l[ (1+2\phi){\rm d}\eta ^2 - 2 \pa_i B {\rm d}x^i
{\rm d}\eta \r. \\
& & - \l. \l[(1-2 \psi )\delta _{ij}+2 \pa_i \pa_j E + h_{ij} \r]
{\rm d}x^i{\rm d}x^j \r]\ , \nn
\er 
where the functions $\phi$, $B$, $\psi$ and $E$ represent the scalar
sector whereas the tensor $h_{ij}$, satisfying $h_i^i = \pa^{i} h_{ij}
= 0$, represent gravitational waves. Note that all these first-order
perturbations are functions of $(\eta, {\bf x})$. For convenience, we
do not write the dependence explicitly.

The tensor perturbations do not couple to the energy density
($\d\rho$) and pressure ($\d p$) inhomogeneities. However, the scalar
perturbations couple to the energy density and pressure which lead to
the growing inhomogeneities. At the linear level, the two types of
perturbations decouple and can be treated separately.

The scalar and tensor perturbations have four and two degrees of
freedom respectively. In the case of tensor perturbations, the two
degrees of freedom correspond to the two polarizations of the
gravitational waves and hence are physical. The scalar perturbations
suffer from the gauge problem.  (For a detailed discussion, see
Refs. \cite{Kodama-Sasa:1984,Mukhanov-Feld:1992}.)  However, it is
possible to construct two gauge invariant variables, which
characterize the perturbations completely, from the metric variables
alone i. e.,
\begin{equation}
\label{eq:gimp}
\Phi \equiv \phi +\frac{1}{a}[(B-E')a]', \qquad
\Psi \equiv \psi - \Hbar (B-E') \, .
\end{equation}
Physically, $\Phi$ corresponds to the Newtonian gravitational potential
and is commonly referred to as the Bardeen potential while $\Psi$ is
related to the perturbations of the $3$-space. For the single
canonical scalar field scenario, we have $\Phi = \Psi$.

$\Phi$ and $\Psi$ are related to the pressure and density
perturbations of a generic perfect fluid {\it via} the perturbed
Einstein's equations. The pressure perturbations, in general, can be
split into adiabatic and entropic (non-adiabatic) parts, by writing
\begin{equation}
\delta p = c_{{\rm s}}^2 \delta\rho + \ov{p}{'} \Gamma \, ,
\label{eq:dPdR}
\end{equation}
where $c_s^2 \equiv {{\ov p}{'}}/{{\ov \rho}{'}}$ is the adiabatic
sound speed
\cite{Wands-Mali:2000,Malik-Wand:2002,Malik-Wand:2004}. The
non-adiabatic part is $\delta p_{\rm nad}\equiv {{\ov
p}{'}}\Gamma$, and
\begin{equation}
\label{defGamma}
\Gamma \equiv \frac{\delta p}{{\ov p}{'}} - 
\frac{\delta\rho}{{\ov \rho}{'}} \,.
\end{equation}
The entropic perturbation $\Gamma$, defined in this way, is
gauge-invariant, and represents the displacement between
hyper-surfaces of uniform pressure and uniform density.  In the
context of canonical single scalar field inflation, only adiabatic
perturbations are present, i. e., $\d p = c_s^2 \d\rho$ (where $c_s^2
= 1$) or $\Gamma = 0$. However, as we shall see in the later sections,
the trans-Planckian inflationary scenario introduces entropic
perturbations as well.

%%%%%%%%%%%%%%%%%%%%% SUBSECTION %%%%%%%%%%%%%%%%%%%%%%%%%%%%%%%%%%

\subsection{Canonical scalar field}

The dominant matter component during inflation is a spatially
homogeneous canonical scalar field $\ov{\vp}(\eta)$ (inflaton). The
Lagrangian density for the canonical scalar field $\vp(\eta,{\ov x})$
propagating in a general curved background is given by
\beq
\label{eq:std-sca}
{\cal L}_{\varphi} =  \frac{1}{2} g^{\mu\nu} \,
\pa_{\mu}\varphi \, \pa_{\nu}\varphi - V(\varphi) \, ,
\eeq
where $V(\varphi)$ is the self-interacting scalar field potential. The
equation of motion and the stress tensor of the scalar field
($\varphi$) in the conformally flat FRW background (\ref{eq:frw}) are
given by
\br
\label{eq:eom-phi}
& & 
{\varphi}'' + 2 \, \Hbar \, {\varphi}' - \nabla^2\varphi + a^2 \,
V_{{,\varphi}}(\varphi) = 0 \, , \\
\label{eq:ST-phi}
& & {T^{\mu}_{\nu}}^{(\vp)} = \pa^{\mu}\vp \, \pa_{\nu}\vp 
- \l[\frac{1}{2} \pa^{\a}\vp \, \pa_{\b}\vp - V(\vp)\r] 
\delta^{\mu}_{\nu} \, ,
\er
where $V_{,\varphi} = \l(dV(\varphi)/d\varphi\r)$ and $\nabla^2$
refers to the Laplacian in the flat space.

Let us consider a small inhomogeneous quantum fluctuations on top of a
homogeneous and isotropic classical background. For the scalar field,
we have
\beq
{\varphi}(\eta, {\bf x}) = \ov{\vp}(\eta) + \delta \varphi
(\eta, {\bf x}) \, ,
\eeq
where one assumes that the perturbation $\delta \varphi$ is small.
The perturbed scalar field ($\delta\vp$) and the perturbed
stress-tensor of the scalar field ($\delta{T^{\mu}_{\nu}}^{(\vp)}$),
like the other scalar-type perturbation functions $(\phi, B, \psi,
E)$, suffer from the gauge problem. (For a detailed discussion, see
Refs. \cite{Kodama-Sasa:1984,Mukhanov-Feld:1992}.)  Similar to
Eq.~(\ref{eq:gimp}), it is possible to define a gauge-invariant
quantity for the perturbed scalar field and the perturbed
stress-tensor, i.~e.,
{\small
\br
\label{eq:gisf}
& & {\delta{\varphi}}^{(gi)} \equiv {\delta{\varphi}} + \ov{\varphi}' (B - E') 
\, , \\
%%%%
& &
\delta{T_{0}^{0}}^{({\rm gi})} \equiv \delta{T_{0}^{0}} + {\ov{T_{0}^{0}}}'
(B - E') \, , \nn \\ 
%%%%
\label{eq:giST}
& &
\delta{T_{i}^{j}}^{({\rm gi})} \equiv \delta{T_{i}^{j}} + {\ov{T_{i}^{j}}}'
(B - E') \, , \\
%%%%%
& & 
\delta{T_{0}^{i}}^{({\rm gi})} \equiv \delta{T_{0}^{j}} + 
\l(\ov{T_{0}^{0}} - \frac{1}{3} \ov{T_{i}^{j}}\r)
(B - E')_{,i} \, . \nn 
\er
}
Separating the homogeneous and perturbed part from
Eq. (\ref{eq:eom-phi}), we have
\br
\label{eq:phi-frw}
& & \ov{\vp}^{''} + 2 \Hbar \ov{\vp}^{'} + a^2 V_{,\vp} = 0 \, , \\
%%%%%
\label{eq:gisfper}
& & {\delta{\varphi}^{(gi)}}'' + 2 \, \Hbar \, {\delta{\varphi}^{(gi)}}' - 
\nabla^2\l({\delta{\varphi^{(gi)}}}\r)   \\ 
& & \qquad \qquad \quad + V_{,\varphi\varphi} \, a^2 \, 
{\delta{\varphi}}^{(gi)} - 4 \ov{\varphi}' \Phi' + 
2 V_{,\varphi} \, a^2 \, \Phi = 0 \nn \, .
\er
Similarly, separating the homogeneous and perturbed part in the
stress-tensor (\ref{eq:ST-phi}), we get
{\small
\br
\label{eq:ST-frw}
& & \!\!\!\!\!\!\!\!\!\!\!\!\!
\ov{{T^0_0}^{(\vp)}} = \frac{1}{2}\l(\frac{\ov{\vp}'^2}{a^2} 
+ V(\ov{\vp})\r); - \ov{{T^i_j}^{(\vp)}} = 
\frac{1}{2}\l(\frac{\ov{\vp}'^2}{a^2}  - V(\ov{\vp})\r) \d^i_j \\
%%%%
& & ^{(gi)}{\delta T^0_0}^{(\vp)} = 
a^{-2} \l[-{\ov{\vp'}}^2 \Phi + {\ov{\vp'}} {\delta\vp^{(gi)}}'
+ V_{,\vp} a^2 \delta \vp^{(gi)}\r]  \, ,\nn \\
\label{eq:per-std}
& & ^{(gi)}{\delta T^0_i}^{(\vp)} = 
a^{-2} {\ov{\vp'}} \delta\vp^{(gi)}_{,i} \, , \\
& & ^{(gi)}{\delta T^i_j}^{(\vp)} = 
a^{-2}\l[{\ov{\vp'}}^2 \Phi - {\ov{\vp'}} {\delta\vp^{(gi)}}'
+ V_{,\vp} a^2 \delta \vp^{(gi)}\r] \delta^{i}_{_j} \, . \nn
\er
}
%

%%%%%%%%%%%%%%%%%%%%% SUBSECTION %%%%%%%%%%%%%%%%%%%%%%%%%%%%%%%%%%

\subsection{Scalar and Tensor Perturbation equations}

In the earlier subsections, we obtained the gauge invariant variables
related to the scalar field $({\delta{\varphi}^{(gi)}}, ^{(gi)}{\delta
T^\mu_\nu}^{(\vp)})$ and metric perturbations $(\Phi, \Psi)$. In this
subsection, we outline the essential steps leading to the scalar and
tensor equations of motion. Even though, this is a standard result,
and can be found in numerous review articles (see, for example, Refs.
\cite{Kodama-Sasa:1984,Mukhanov-Feld:1992}), we have given here the key
steps for future reference.

\indent From Eq. (\ref{eq:per-std}), it is easy to see that the non-diagonal
space-space components of the stress-tensor are absent. This leads to
the condition that $\Phi = \Psi$. Thus, the Einstein's equations for
the perturbed FRW metric (\ref{eq:per-frw}) in terms of the gauge
invariant quantities are:
\begin{subequations}
\label{eq:hydro-eq}
\br
\label{eq:hydro-eq1}
\nabla^2 \Phi - 3 \Hbar \Phi' - 3 \Hbar^2 \Phi = 
\frac{1}{2 \, \MPl}\, a^2 \,\delta {T^0_0}^{({\rm gi})} & & \\
%%%%
\label{eq:hydro-eq2}
\frac{(a \Phi)'_{,i}}{a} = \frac{1}{2 \, \MPl}\, a^2 \,
\delta {T^0_i}^{({\rm gi})} & & \\
%%%%
\label{eq:hydro-eq3}
\Phi'' + 3 \Hbar \Phi' + (2 \Hbar' + \Hbar^2) \Phi = 
\frac{1}{2 \, \MPl}\, a^2 \, \delta {T^i_i}^{({\rm gi})} \, , & &  
\er
\end{subequations}
where $\delta {T^\mu_\nu}^{({\rm gi})}$ is given by
Eq. (\ref{eq:per-std}). The three perturbed Einstein's equations can
be combined to form a single differential equation in $\Phi$:
{\small
\beq
\label{eq:Std-Phi}
\Phi^{''} - \nabla^{2}\Phi 
+ 2 \l(\Hbar - \frac{\ov{\vp}^{''}}{\ov{\vp}^{'}}\r) \Phi^{'} 
+ 2 \l(\Hbar^{'} - \Hbar \frac{\ov{\vp}^{''}}{\ov{\vp}^{'}}\r) \Phi = 0 \, .
\eeq
}
The system of perturbation equations (\ref{eq:Std-Phi},
\ref{eq:hydro-eq2}, \ref{eq:gisfper}) is quite complex. In order 
to extract the physical content more transparent, these equations are
expressed in terms of two new variables -- $Q$ and $u$ -- which are
linearly related to $\Phi$ and $\d\vp^{(gi)}$, i. e.,
\beq
\label{eq:MS-u-def}
Q = a\left[\delta \vp + \ov{\vp}'\frac{\psi }{\Hbar}\right]~;~
u = \frac{a}{\ov{\vp}'} \Phi \, .
\eeq
$Q$ is a gauge-invariant (Mukhanov-Sasaki)
\cite{Kodama-Sasa:1984,Mukhanov-Feld:1992} variable whose equation of
motion is homogeneous. This ensures that one can quantize $Q$ in the
standard way using the Lagrangian associated with its equation of
motion.  At the early stages of inflation where the quantum effects
are important, the equation of motion of $Q$ helps in quantizing the
fields and fixing the initial conditions. However, at the end stages
of inflation where the relevant modes have crossed the Hubble radius
and behave classically, it is easier to analyze the equation of motion
of $u$.

The equation of motion of $Q$ is derived as follows: Substituting
$\d\vp$ in terms of $Q$ in Eq.~(\ref{eq:gisfper}), and using the
relations (\ref{eq:Std-Phi}, \ref{eq:phi-frw}, \ref{eq:hydro-eq2}), we
get:
\beq
\label{eq:MSeq-Std}
Q'' -  \, \nabla^2 Q - \frac{z''}{z} Q = 0 \, ,
\eeq
where 
\beq
\label{eq:def-z}
z = \frac{a \, (\Hbar^2 - \Hbar')^{1/2}}{\Hbar \, c_s}
= a \frac{\bar{\varphi}'}{\Hbar} \, .
\eeq

The equation of motion of $u$ is obtained by substituting the 
transformation (\ref{eq:MS-u-def}) in Eq. (\ref{eq:Std-Phi}):
\beq
\label{eq:ueq-Std}
u'' -  \, \nabla^2 u - \frac{\theta''}{\theta} u = 0 \, , 
\eeq
where 
\beq
\theta = \frac{\Hbar}{a} \l[\frac{2}{3}(\Hbar^2 - \Hbar')\r]^{-1/2} 
= \frac{\Hbar}{a \bar{\varphi}'} \, .
\eeq
$Q$ is related to another gauge-invariant quantity $\zeta (\equiv -
(\Hbar/{\rho^{'}}) \delta\rho + \psi)$ by the relation $Q = 2 c_s\,z
\zeta$. The quantity $\zeta$ is time-independent on scales larger than
the Hubble radius and, more importantly, related to the large scale 
CMB anisotropies (via the Sachs-Wolfe effect) \cite{Liddle-Lyth:2000-bk}. 

Decomposing $Q$ into Fourier space, we have
\beq
\label{eq:SPer0}
\mu_{_S}'' +  \l[k^2  - \frac{z''}{z} \r] \mu_{_S} = 0 \, ,
\eeq
where $k = |{\bf k}|$ and $\mu_{_S} = - Q_k = - 2 c_s\,z \zeta$. The
above equation is similar to a time-independent Schr\"odinger
equation
 where the usual role of the radial coordinate is now played
by the
 conformal time whose effective potential is $U_{_{\rm
S}}\equiv
 {z''}/{z}$ \cite{Wang-Mukh:1997,Martin-Schw:2002}. The
scalar perturbation spectrum per logarithmic interval
 can then be
written in terms of the modes $\mu_{_S}$ as
\beq 
\left[k^3\; {\cal P}_{S}(k)\right]
=\left(\frac{k^3}{2\pi^2}\right)\, \left(\frac{\vert
\mu_{_S}\vert}{z}\right)^2 \, ,
\label{eq:pS0}
\eeq 
and the expression on the right hand side is to be evaluated when 
the physical wavelength $(k/a)^{-1}$ of the mode corresponding to 
the wavenumber ${\bf k}$ equals the Hubble radius $H^{-1}$. 

Before proceeding to the next section, we obtain the tensor
perturbation equation in the FRW background. As we mentioned earlier,
the tensor perturbations $h_{ij}(\eta, {\bf x})$ do not couple to the
energy density. These represent free gravitational waves and satisfy
the equation:
\begin{equation}
\label{eq:TPer0}
\mu_{_T}^{''} + \bigl( k^2 - \frac{a''}{a} \bigr) \mu_{_T} \, = \, 0 \, .
\end{equation}
where $\mu_T \, \equiv \, a \, h_k $. This equation is very similar to
the corresponding equation (\ref{eq:SPer0}) for scalar gravitational
inhomogeneities, except that in the effective potential $(U_{_{\rm
T}}\equiv {a''}/{a})$ the scale factor $a(\eta)$ is replaced by
$z(\eta)$.

%%%%%%%%%%%%%%%%%%%%%%%%%%%%%%%%%%%%%%%%%%%%%%%%%%%%%%%%%%%%%%%%%%%%%%%%%%
%%%%%%%%%%%%%%%%%%%%%%%%%  NEW SECTION %%%%%%%%%%%%%%%%%%%%%%%%%%%%%%%%%%%
%%%%%%%%%%%%%%%%%%%%%%%%%%%%%%%%%%%%%%%%%%%%%%%%%%%%%%%%%%%%%%%%%%%%%%%%%%

\section{Modified Dispersion relation Lagrangian}
\label{sec:MDR}
 
In this section, we briefly discuss the general covariant formulation
describing a scalar field with modified dispersion relation and derive
the corresponding stress-tensor.

As discussed in the introduction, to keep the calculations tractable,
we will assume that the scalar field with the high frequency
dispersion relation is of the form 
\begin{equation}
\omega^2 = |\vec{k}|^2 + b_{_{11}} |\vec{k}|^4 \, ,
\end{equation}
where $b_{_{11}} > 0$. The above dispersion relation breaks the local
Lorentz invariance explicitly while it preserves rotational and
translational invariance.  Even though, the modified dispersion
relation breaks the local Lorentz invariance explicitly, it was shown
in Ref. \cite{Jacobson-Matt:2000b} that a covariant formulation of the
corresponding theory can be carried out by introducing a unit
time-like vector field $u^{\mu }$ which defines a preferred rest
frame.

%%%%%%%%%%%%%%%%%%%%% SUBSECTION %%%%%%%%%%%%%%%%%%%%%%%%%%%%%%%%%%

\subsection{Covariant Lagrangian}
\label{sec:Cov-Lag}

The action for a scalar field with the modified dispersion relation
takes the form~\cite{Jacobson-Matt:2000b,Lemoine-Lubo:2001}
\br
\label{eq:mdr-sca}
S &=&\int{\rm d}^4x\sqrt{-g}~({\cal L}_{\varphi}+{\cal L}_{_{\rm
cor}}+{\cal L}_u),
\er
where ${\cal L}_{\varphi}$ is the standard Lagrangian of a minimally
coupled scalar field given by Eq. (\ref{eq:std-sca}). The last two
terms --- ${\cal L}_{_{\rm cor}}$ and ${\cal L}_u$ --- contribute to
the modified dispersion relation of the scalar field.  ${\cal
L}_{_{\rm cor}}$ corresponds to the non-linear part of the dispersion
relation while ${\cal L}_u$ describes the dynamics of the vector field
$u^\mu$. The two corrective Lagrangians have the form
\begin{subequations}
\br
\label{eq:lcor}
{\cal L}_{_{\rm cor}}&=& -  b_{11}
\left(\DD^{2}\varphi\right)^2,  \\
\label{eq:lu}
{\cal L}_u&=& - \lambda(g^{\mu\nu}u_\mu u_\nu - 1)-
d_1F^{\mu \nu}F_{\mu \nu} \, ,
\er
\end{subequations}
where 
\br
\label{eq:def-Fmn}
F_{\mu \nu }&\equiv& \nabla _{\mu }\,u_{\nu}-\nabla_\nu\,u_\mu \, , \\
\label{eq:DDvp}
{\DD}^2 \varphi 
&=& \perp^{\alpha\beta}\nabla_\alpha\nabla_\beta\varphi
+u^\alpha\nabla_\alpha\varphi\nabla_\beta u^\beta \, , \\
\label{eq:def-per}
\perp_{\mu\nu} &\equiv& - g_{\mu\nu}+ u_{\mu} u_{\nu}  \quad .
\er
The covariant derivative associated with the metric $g_{\mu \nu}$ is
$\nabla_{\mu }$ while $b_{11}$ and $d_1$ are arbitrary (dimensional)
constants.  The tensor $\perp_{\mu\nu}$ gives the metric on a slice of
fixed time while ${\DD}^2$ is proportional to the Laplacian operator
on the same surface. The fact that $u^\mu$ is a unit time-like vector
($u^\mu u_{\mu} = 1$) is enforced by the Lagrange multiplier
$\lambda$. In the above, $b_{11}, d_1$ have the dimensions of inverse
mass square and mass square, respectively while $u_{\mu}$ is
dimensionless.

The equation of motion for $\vp$ and $u_{\mu}$ obtained by varying the
action (\ref{eq:mdr-sca}), respectively, are
{\small
\br
\label{eq:DE-vp}
 \nabla^{\mu}\nabla_{\mu}\vp + V_{,\vp} &=& 2 b_{_{11}} 
\l[\nabla_{\mu}\l(\DD^2\vp \, u^{\mu} \, \nabla_{\nu}u^{\nu}\r) \r. \\
%%%%%
& & \qquad \qquad \qquad  \l.
- \nabla_{\mu}\nabla_{\nu}\l(\DD^2\vp \, \perp^{\mu\nu}\r) \r] \, , \nn \\
%%%%%
\label{eq:DE-u}
2 d_{1} \nabla_{\nu} F^{\nu\mu} - \lambda u^{\mu} 
&=& - b_{_{11}} \l[\nabla^{\mu}\l(\DD^2\vp u^{\nu} \pa_{\nu}\vp\r) \r. \\
%%%%%
& -& \l. \nabla^{\mu}\nabla_{\nu}\vp \, u^{\nu} \DD^2\vp 
- \nabla_{\nu}\l(\nabla^{\mu}\vp u^{\nu}\r) \DD^2\vp \r] \, . \nn
\er
}

\noindent From the above equation, we get
\br
\label{eq:def-lam}
\lambda &=& 
b_{_{11}} u_{\mu} \l[\nabla^{\mu}\l(\DD^2\vp u^{\nu} \pa_{\nu}\vp\r)
- \nabla^{\mu}\nabla_{\nu}\vp \, u^{\nu} \DD^2\vp \r. \\
& & \l. - \nabla_{\nu}\l(\nabla^{\mu}\vp u^{\nu}\r) \DD^2\vp \r] 
+ 2 d_1 u_{\mu} \nabla_{\nu} F^{\nu\mu} \, .\nn
\er
Using the results of Appendix (\ref{sec:FRW-bkg}), it is easy to show
that for the FRW background $\ov{\vp}$ satisfies
Eq. (\ref{eq:phi-frw}) which is same as that of the canonical scalar
field. The field equation for $u_{\mu}$ gives $\ov{\lambda} = 0$.

Before we proceed to the computation of the stress-tensor, it is
important to know the current astrophysical constraints on the
parameters of the model \cite{Mattingly:2005}: The constraints of the
parameter $d_1$ comes from the big-bang nucleosynthesis
\cite{Carroll-Lim:2004} and the solar system tests of general relativity
\cite{Graesser-Jenk:2005}.  These give:
\beq
0 < \frac{d_1}{\MPl} < \frac{1}{7} \, .
\eeq
The constraints on the parameter $b_{11}$ comes from the observations
of highest energy cosmic rays \cite{Gagnon-Moor:2004}.  Using
effective field theory with higher-dimensional operators, resulting in
the modified field theory with a dispersion relation, it was shown
that for various standard model particles
\beq
b_{11} \, \MPl < 5 \times 10^{-5} \, .
\eeq
% 

%%%%%%%%%%%%%%%%%%%%% SUBSECTION %%%%%%%%%%%%%%%%%%%%%%%%%%%%%%%%%%

\subsection{Stress tensor}
\label{sec:Str-Ten}

In this subsection, we obtain the stress-tensor for the scalar field
defined in Eq. (\ref{eq:mdr-sca}). Formally, the stress-tensor for a
general Lagrangian containing up to first order derivative in the
metric is given by (cf. Ref. \cite{Landau-Lifs:1976-bk2}, p. 272)
\beq
\label{ll}
T_{\mu\nu}=-g_{\mu\nu}{\cal L} + 2 
\frac{\partial{\cal L}}{\partial g^{\mu\nu}}
- \frac{2}{\sqrt{-g}}\partial_\rho\left(\!\sqrt{-g}
\frac{\partial{\cal L}}{\partial(\partial_\rho g^{\mu\nu})}\right).
\eeq
For simplicity, we will separate the contributions from the three 
different Lagrangians defined in Eq. (\ref{eq:mdr-sca}), i. e.,
\be
T_{\mu\nu} = T^{(\varphi)}_{\mu\nu} +
             T^{(u)}_{\mu\nu} +
	     T^{(cor)}_{\mu\nu} \, .
\ee
The stress-tensor $T_{\mu \nu}^{^{(\varphi)}}$ corresponding to the
canonical scalar field Lagrangian is given in Eq. (\ref{eq:ST-phi}).
The stress-tensor corresponding to the Lagrangian ${\cal L}_{_{u}}$
can easily be obtained and is given by
\begin{eqnarray}
T_{\mu \nu}^{^{(u)}} &=& 
d_1 g_{\mu \nu} F_{\ve\vk} F^{\ve\vk} - 
4 d_1 F_{\mu \ve} F_{\nu \vk} g^{\ve\vk} \nn \\
& & + \lambda \l[g_{\mu \nu} 
(g_{\ve\vk} u^{\ve} u^{\vk} -1) - 2 u_\mu u_\nu \r] \, .
\label{eq:ST-u}
\end{eqnarray}
However, the stress-tensor corresponding to ${\cal L}_{_{cor}}$ is
much more involved. For the sake of continuity we give below only the
final result while the steps leading to the result is given in
Appendix (\ref{sec:FRW-bkg}).  We get,
{\small
\beq
\label{eq:ST-cor}
T_{\mu\nu}^{^{\rm (cor)}} =  b_{_{11}} g_{_{\mu \nu}} ({\DD}^2 \varphi)^2  
+ b_{_{11}} E_{_{\mu\nu}} \DD^2\varphi 
+ 4 b_{_{11}} C_{_{\mu \nu}}^{\rho \vk}  \partial_{\vk}\varphi  \, 
\partial_\rho[{\DD}^2 \varphi] \, ,
\eeq
}
\noindent where,
\br
E_{\mu\nu}& =& 4 \l[
\partial_\rho[\ln(\sqrt{-g})] \, C_{\mu \nu}^{\rho \vk}\, 
\partial_{\vk} \varphi 
+ \partial_\rho \left(C_{\mu \nu}^{\rho \vk} \partial_{\vk} \varphi\r)\r.\nn \\
\label{eq:Def-Emn}
& & - \l.  A_{\mu \nu}^{\ve\vk} \partial_{\ve} \partial_{\vk}
\varphi - B^{\vk}_{\mu\nu} \partial_{\vk}  \varphi \r] \\
%%%%%
C^{\rho \vk}_{\m\n} &=& \frac{1}{2} \l[ g_{\m\n} g^{\vk \rho} - 
\delta^\rho_\m \delta^\vk_\m - \delta^\vk_\m \delta^\rho_\n  
+ u_{_\m} \delta^\rho_\n u^{\vk}  - u_{_\m} u_{_\n} g^{\vk \rho} \r. \nn\\
\label{eq:Def-C}
 &+& \l. u_{_\m} \delta^\vk_\n u^{\rho} + u_{_\n} \delta^\vk_\m u^{\rho} 
- g_{\m\n} u^{\rho} u^{\vk} + \delta^\rho_\m u_\n u^{\vk}  \r] \, .
\er
Before proceeding with the evaluation of the perturbed stress-tensor
we would like to mention the following point: Using the results of 
Appendix (\ref{sec:FRW-bkg}), it is clear that $T_{\mu\nu}^{(u)}$ and 
$T_{\mu\nu}^{(cor)}$ vanishes in the unperturbed FRW background. 
Hence, the equations determining the evolution of the scale 
factor, i. e.,
\beq
3 \Hbar^2 = \frac{a^2}{\MPl} \ov{T^0_0}^{(\vp)}~;~
2 \Hbar^{'} + \Hbar^2 = \frac{a^2}{3 \MPl} \ov{T^i_i}^{(\vp)} \, ,
\eeq
remain the same as in the canonical scalar field inflation. It is also
worth mentioning that the trans-Planckian corrections do not play any
role on the expansion of the FRW background while, as we will see in
the next section, the trans-Planckian corrections affect the metric
and inflaton perturbations.

In this work, we will focus on the power-law inflation, for which, the
scale factor is given by
\beq
\label{eq:polaw}
a(t) = \l(a_0 \, t^p\r)\quad {\rm or} \quad  
a(\eta) = \l(\frac{-\eta}{-\eta_0} \r)^{(\beta+1)}
\, ,
\eeq 
where $p > 1$, $\beta \leq -2$ ($\beta = -2$ corresponds to de
Sitter), $a_0$ is a constant,
\beq
\beta = -\l(\frac{2p-1}{p - 1}\r)
\quad{\rm and}\quad
(- \eta_0) = \frac{a_0^{-1/p}}{(p - 1)} \, .
\eeq
The scalar field potential and other background field parameters 
are given by ($q = \sqrt{2/p}$)
\br
V = v_o M_{_{\rm Pl}}^4 
\exp \l(- q \frac{\ov{\varphi}}{M_{_{\rm
Pl}}}\r) &;& \Hbar = \frac{-(1 + \beta)}{(-\eta)}~;\\
%%%%%%
{\ov \vp} = \sigma_o \MPlh \ln\l(\frac{-\eta}{-\eta_0}\r) 
&;&\sigma_o = \sqrt{2 \beta (\beta +1)}~; \nn \\ 
%%%%
\label{eq:PLaw-para}
a_0 = - \frac{\sqrt{(\beta + 1)(1+2 \beta)}}{\sqrt{v_o} \eta_0 M_{_{\rm
Pl}}}  &;& 
\frac{z''}{z} = \frac{a''}{a} = \frac{\beta (\beta +1)}{(-\eta)^2} \, . \nn
\er
%

%%%%%%%%%%%%%%%%%%%%%%%%%%%%%%%%%%%%%%%%%%%%%%%%%%%%%%%%%%%%%%%%%%%%%%%%%%
%%%%%%%%%%%%%%%%%%%%%%%%%  NEW SECTION %%%%%%%%%%%%%%%%%%%%%%%%%%%%%%%%%%%
%%%%%%%%%%%%%%%%%%%%%%%%%%%%%%%%%%%%%%%%%%%%%%%%%%%%%%%%%%%%%%%%%%%%%%%%%%

\section{Perturbed stress tensor}
\label{sec:per-st}

In this section, we will obtain the perturbed stress-tensor for the
scalar field with modified dispersion relation (\ref{eq:mdr-sca}) in
the perturbed FRW background (\ref{eq:per-frw}).  As in the previous
section, we will separate the contributions to the perturbed
stress-tensor from the three different Lagrangians defined in
Eq. (\ref{eq:mdr-sca}), i. e.\footnote{The mixed stress-tensor $\delta
T^{\mu}_{\nu}$ is given by
\beq
\delta T^{\mu}_{\nu} \equiv \delta \l(g^{\mu\ep} T_{\ep\nu}\r) 
= \ov{g^{\mu\ep}} (\d T_{\ep\nu}) + (\d g^{\mu\ep}) \ov{T_{\ep\nu}} 
\label{eq:rel-sts}
\eeq
},
\beq
\delta T^{\mu}_{\nu} = \delta {T^{\mu}_{\nu}}^{(\varphi)} +
                       \delta {T^{\mu}_{\nu}}^{(u)} +
		       \delta {T^{\mu}_{\nu}}^{(cor)} \, .
\label{eq:per-stmn}
\eeq
The first term in the RHS of the above expression corresponds to the
perturbed stress-tensor of the canonical scalar field Lagrangian and
is given by Eqs. (\ref{eq:per-std}). In the rest of the section, we
will obtain contributions from the other two terms.

Perturbing Eq. (\ref{eq:ST-u}) and using the fact that $F_{\mu\nu}$
vanishes for the FRW background [See Appendix (\ref{sec:FRW-bkg})], we
get
\beq
\label{eq:per-stu}
\delta {T^{\mu}_{\nu}}^{(u)} = - 2 \, \delta^\mu_0 \, 
\delta^0_\nu \, (\delta\lambda) \, ,
\eeq
where $\d\lambda$ is given by Eq. (\ref{eq:p-lam}). Using the fact
that $\ov{{\DD}^2(\varphi)}$ and $\ov{\pa_{\rho}{\DD}^2(\varphi)}$
vanish for the FRW background [See Appendix (\ref{sec:FRW-bkg})],
the perturbation of Eq. (\ref{eq:ST-cor}) takes the following simple
form:
\beq
\d T^{(cor)}_{\mu \nu} = 
4 b_{11} \l[ \ov{E_{\mu\nu}}\, \d(\DD^2 \varphi) + 
\, \ov{C^{\rho 0}_{\mu\nu}} \, \ov{\varphi}' \, 
\pa_{\rho}(\d\DD^2\varphi)  \r] \, .
\eeq
Substituting the relation (\ref{eq:d-DD}) for $\d(\DD^2 \varphi)$ and
Eqs. (\ref{eq:bABCE1}) in the above expression, we get
\br
\label{eq:per-stcor}
& & \!\!\!\!\!\!\!\!\!\!\!\! 
\delta {T^{\mu}_{\nu}}^{(cor)} = \frac{2 b_{11}}{a^4} 
\l[ 5 \frac{{\cal H}}{a} {\ov{\vp}^{'}}^2 \nabla^2 \xi^{(gi)}  
- \frac{1}{a} {\ov{\vp}'}^2 \nabla^2 {\xi^{(gi)}}' \r. \\
%%%%
& & \quad ~~ -  \l. \l({\ov{\vp}^{''}} + 4 {\cal H} {\ov{\vp}^{'}} \r) 
\nabla^2(\d\vp) + {\ov{\vp}^{'}} \nabla^2(\d\vp)' \r] 
\d^{\mu}_{_{0}}\d^{0}_{_{\nu}} \, ,\nn 
\er
where $\xi$ is defined in Eq.~(\ref{eq:def-xi}).  Substituting
Eqs. (\ref{eq:per-std}, \ref{eq:per-stu}, \ref{eq:per-stcor}) in
Eq. (\ref{eq:per-stmn}), we obtain the perturbed gauge-invariant 
stress-tensor to be:
\begin{subequations}
\label{eq:per-fin}
\br
{\delta T^0_0}^{(gi)}&=& \frac{1}{a^{2}}
\l[-{\ov{\vp'}}^2 \Phi + {\ov{\vp'}} {\delta\vp^{(gi)}}^{'}
+ V_{,\vp} a^2 \delta \vp^{(gi)}\r]  \nn \\
%%%%
\label{eq:per-fin1}
& & \qquad + \, 
\frac{4 d_{_1}}{a^2} \l[\nabla^2 \Phi - \frac{1}{a} \nabla^2 {\xi^{(gi)}}^{'}
\r] \, , \\
%%%%
\label{eq:per-fin2}
{\delta T^i_j}^{(gi)} &=& \l[{\ov{\vp'}}^2 \Phi - {\ov{\vp'}} \delta\vp'^{(gi)}
+ V_{,\vp} a^2 \delta \vp^{(gi)}\r] \delta^{i}_{_j} \, , \\   
%%%%%
\label{eq:per-fin3}
{\delta T^0_i}^{(gi)}&=& a^{-2} {\ov{\vp'}} \delta\vp_{,i}^{(gi)} \, .
\er
\end{subequations}
Following points are worth-noting regarding the above result: Firstly,
the two corrective Lagrangians -- ${\cal L}_{u}$ and ${\cal
L}_{_{cor}}$ -- contributes only to the perturbed energy density
$(\d\rho)$. This implies that the non-diagonal space-space components
of the stress-tensor are absent leading to the condition that $\Phi =
\Psi$. This also implies that the constraint equation
(\ref{eq:hydro-eq2}) remains unchanged even for trans-Planckian
inflation. Secondly, since the trans-Planckian corrections do not
change the pressure perturbations, the perturbation equations for the
tensor modes do not change. Hence, the tensor perturbation equations
remain unchanged. Recently, Lim \cite{Lim:2004} had show that general
Lorentz violating models (with out taking into account the higher
derivatives of the scalar field) can modify the pressure perturbations
and hence the tensor perturbation equations. However, in our specific
Lorentz violating model, this is not the case. { This indicates that
the well-know consistency relation between the scalar and tensor ratio
will also be broken in this model \cite{Hui-Kinn:2001,Ashoorioon-Mann:2004,Ashoorioon-Hovd:2005}.}

Thirdly, it is
interesting to note that the trans-Planckian contributions to the
energy density go as $k^2$. Hence as one would expect, in the
super-Hubble scales, {\it only} the canonical scalar field contributes
significantly in these scales. Lastly, these can have two significant
implications on the perturbation spectrum: (a) The speed of
propagation of the perturbations ($c_s^2$) can be different from that
of the standard single-scalar field inflation. In the case of single
scalar field inflation, we know that $c_s^2 = 1$.  However, due to the
extra contributions to the energy density, this can no longer be
true. (b) The perturbations need not be purely adiabatic. $\xi$ can
act as an extra scalar field during the inflation and hence can act as
a source. This can introduce non-adiabatic (entropic) perturbations.
(See, for example, Ref. \cite{GrootNibbelink-Vant:2001,vanTent:2003})
We will discuss more on these in the following sections.

%%%%%%%%%%%%%%%%%%%%%%%%%%%%%%%%%%%%%%%%%%%%%%%%%%%%%%%%%%%%%%%%%%%%%%%%%%
%%%%%%%%%%%%%%%%%%%%%%%%%  NEW SECTION %%%%%%%%%%%%%%%%%%%%%%%%%%%%%%%%%%%
%%%%%%%%%%%%%%%%%%%%%%%%%%%%%%%%%%%%%%%%%%%%%%%%%%%%%%%%%%%%%%%%%%%%%%%%%%

\section{Scalar Perturbation equation}
\label{sec:Sca-Per-Eq}

Substituting Eqs. (\ref{eq:per-fin}) in (\ref{eq:hydro-eq}), the
first-order perturbed Einstein's equations take the following form,
\begin{subequations}
\label{eq:Per-EinTP}
{\small
\br
& & \nabla^2 \Phi - 3 \Hbar \Phi' - 3 \Hbar^2 \Phi = 
\frac{2 d_{_1}}{\MPl} \l[\nabla^2 \Phi - \frac{1}{a} 
\nabla^2 {\xi^{(gi)}}^{'}\r] \\
& &  \qquad \qquad \quad
+ \, \frac{1}{2 \MPl} \l[-{\ov{\vp'}}^2 \Phi + {\ov{\vp'}} 
{\delta\vp^{(gi)}}^{'} + V_{,\vp} a^2 \delta \vp^{(gi)}\r] \, , \nn \\
%%%%
& & \Hbar \Phi + \Phi' = \frac{1}{2 \MPl}\, 
{\ov{\vp'}} \delta\vp^{(gi)} \, , \\
%%%%
& & \Phi'' + 3 \Hbar \Phi' + (2 \Hbar' + \Hbar^2) \Phi = 
\frac{1}{2 \MPl}\,\l[{\ov{\vp'}}^2 \Phi - {\ov{\vp'}} \delta\vp'^{(gi)}
\r. \nn \\
& & \l. \qquad \qquad \qquad \qquad \qquad \qquad \quad 
+~~ V_{,\vp} a^2 \delta \vp^{(gi)}\r] \, .
\er
}
\end{subequations}
As in the canonical scalar field inflation, the three perturbed
Einstein's (\ref{eq:Per-EinTP}) can be combined to give 
\begin{multline}
\label{eq:MDR-Phi}
\Phi^{''} - \l(1 - \frac{2 d_{_1}}{\MPl}\r) \nabla^{2}\Phi
+ 2 \l(\Hbar - \frac{\ov{\vp}^{''}}{\ov{\vp}^{'}}\r) \Phi^{'}  \\
+ 2 \l(\Hbar^{'} - \Hbar \frac{\ov{\vp}^{''}}{\ov{\vp}^{'}}\r) \Phi
 =  \frac{2 d_{_1}}{\MPl} \frac{1}{a} \nabla^2 {\xi^{(gi)}}^{'} \, . 
\end{multline}
Perturbing the field equations (\ref{eq:DE-vp}, \ref{eq:DE-u}), we get
{\small
\br
\label{eq:per-vpTP}
& & \!\!\!\!\! 
{\delta{\varphi}^{(gi)}}'' + 2 \, \Hbar \, {\delta{\varphi}^{(gi)}}' - 
\nabla^2\l({\delta{\varphi^{(gi)}}}\r) + V_{,\varphi\varphi} \, a^2 \, 
{\delta{\varphi}}^{(gi)}  \\ 
%%%%
& & - 4 \ov{\varphi}' \Phi' + 2 V_{,\varphi} \, a^2 \, \Phi 
+ \frac{2 b_{_{11}}}{a^2}\l[\nabla^4 \d\vp^{(gi)} - 
\frac{\ov{\vp}'}{a} \nabla^4 \xi^{(gi)} \r] = 0 \nn \, , \\
%%%%
& & \!\!\!\!\! 
\pa_m \l[\l(1 - \frac{c_1}{\MPl} \frac{\nabla^2}{a^2}\r)\Phi - 
2 \Hbar \l(1 + \frac{c_1}{2 \MPl}\frac{\nabla^2}{a^2}\r)
\Phi'\r] \nn \\
%%%%
\label{eq:per-uTP}
& & - \frac{1}{a} \pa_m \l[{\xi^{(gi)}}^{''} - 3 \Hbar {\xi^{(gi)}}^{'} - 
\frac{b_{11}}{2 d_1} \frac{\ov{\vp}^2}{a^2} \nabla^2\xi^{(gi)}\r] = 0 \, ,
\er
}
where $c_1 = (M_{_{\rm Pl}}^4 b_{11})/d_1$ is a dimensionless
constant.

Following points are interesting to note regarding the above results:
(i) The spatial higher derivatives appear {\it only} in the equation
of motion of $\d\vp$ and not in metric perturbation equation $\Phi$.
Unlike the standard inflation, the perturbations are not purely
adiabatic and the speed of propagation of the perturbations is less
than unity i. e. $c_s^2 = 1 - 2 d_1/\MPl$.  (ii) In the case of
canonical scalar-field inflation, the two dynamical variables $\Phi$
and $\d\vp^{(gi)}$ are related by the constraint equation
(\ref{eq:hydro-eq2}).  In our model, $\Phi$ and $\d\vp^{(gi)}$ are
again related by the same constraint equation, however $\xi$ is
related to the other fields {\it via} the equations of motion.  Hence,
unlike the standard inflation, we have two sets of independent
variables. (iii) In the case of canonical scalar-field inflation,
$\Phi$ and $\d\vp^{(gi)}$ can be combined into a single variable ---
Mukhanov-Sasaki ($Q$) variable --- in terms of which we can obtain the
perturbation spectrum. However, in this model, as in the case of
multi-field inflation models \cite{GrootNibbelink-Vant:2001}, the
equations of motion in terms of the Mukhanov-Sasaki variables are
coupled. In the rest of the section, we derive the equation of motion
of the Mukhanov-Sasaki variables corresponding to $\d\vp^{(gi)}$ and
$\xi$.

Substituting $\d\vp$ in-terms of $Q$ in Eq. (\ref{eq:per-vpTP}), and
using the relations (\ref{eq:MDR-Phi}, \ref{eq:phi-frw},
\ref{eq:hydro-eq2}), we get
\br
Q'' -  \l(1 - \frac{2 b_{11}}{a^2}\nabla^2 \r) \nabla^2Q - 
\frac{z''}{z} Q =  \frac{2 d_1}{\MPl} \nabla^2 {\cal S}(\eta) \, ,
\er
where
\beq
{\cal S} =   \ov{\vp}^{'} \l[
\frac{Q_{\xi}^{(2)}}{\Hbar} +  \frac{c_1}{a^{1/2}} \nabla^2 Q_{\xi}^{(1)} 
\r] \, ,
\eeq
and $Q_{\xi}^{(1)}, Q_{\xi}^{(2)}$ are the gauge-invariant variables
associated with $\xi$ and are given by:
\beq
Q_{\xi}^{(1)} = a^{-3/2} \l[ \xi + \frac{a}{\Hbar} \psi\r]\, ; \,
Q_{\xi}^{(2)} = \xi^{'} - a \phi \, .
\eeq
Substituting for $\xi$ in-terms of $Q_{\xi}^{(1)}$ in (\ref{eq:per-uTP}), 
we get,
{\small
\br
& & \!\!\!\!\!\!\!\!\!
\Bigg[ a^{3/2} \l({Q_{\xi}^{(1)}}^{''} + \l[\frac{3}{2}\Hbar^{'} 
- \frac{9}{4}\Hbar^{2} - 
\frac{b_{_{11}} {\ov \phi^{'}}^2}{2 d_1 a^2} \nabla^2\r]{Q_{\xi}^{(1)}} 
\r) \\
%%%%
& & \!\!\!\!\!\!\!\!\! - \frac{{\ov \phi^{'}}}{\MPl \Hbar}
\l(\frac{{\ov \phi^{''}}}{{\ov \phi^{'}}} - 
\frac{\Hbar^{'}}{\Hbar} - \Hbar - \frac{c_1 \Hbar}{2 \MPl a^2} \nabla^2\r)
Q - \frac{a}{\Hbar} \nabla^2{\Phi}\Bigg ],m  = 0 \, .\nn 
\er
}
Decomposing $Q, Q_{\xi}^{(1)}, Q_{\xi}^{(2)}$ into Fourier space, we have
{\small
\br
\label{eq:SPerTP}
& & \!\!\!\!\! 
\mu_{_S}^{''} + \l[ k^2 + \frac{2 b_{_{11}}}{a^2(\eta)} k^4 -
\frac{z^{''}}{z} \r] \mu_{_S} =  
- \frac{2 d_1}{\MPl} k^2 {\cal S}_{k}(\eta) \\
%%%%
\label{eq:UPerTP}
& & \!\!\!\!\! \mu_{_\xi}^{''} + \l[\frac{3}{2}\Hbar^{'} 
- \frac{9}{4}\Hbar^{2} + 
\frac{b_{_{11}} {\ov \phi^{'}}^2}{2 d_1 a^2} k^2 \r]{\mu_{_\xi}} 
= a^{-3/2} \l[\frac{a k^2}{\Hbar}\Phi_k  \r.
\\
%%%%
& & \l. \qquad \qquad + \frac{{\ov \phi^{'}}}{\MPl \Hbar} 
\l(\frac{{\ov \phi^{''}}}{{\ov \phi^{'}}} - 
\frac{\Hbar^{'}}{\Hbar} - \Hbar + \frac{c_1 \Hbar}{2 \MPl a^2} k^2\r)
\mu_{_S} \r] \, , \nn 
\er
}
\noindent 
where the Fourier transform of $Q$, $Q_{\xi}^{(1)}$, $Q_{\xi}^{(2)}$,
respectively, are $\mu_{_S}, \mu_{_\xi}, Q_{k}^{(2)}$ and
\beq
\label{eq:sou-fin}
{\cal S}_k(\eta) = \ov{\vp}^{'} \l[
\frac{Q_{k}^{(2)}}{\Hbar} - \frac{c_1}{a^{1/2}} k^2 \mu_{_\xi} \r] \, .
\eeq

Eqs. (\ref{eq:SPerTP}, \ref{eq:UPerTP}) are the main results of our
paper, regarding which we would like to stress the following points:
Firstly, in the earlier analyses, the equation of motion of the
Mukhanov-Sasaki variable ($Q$) was assumed to be satisfy the
differential equation Eq. (\ref{eq:SPerTP}) in which the source term
was assumed to be zero. We have shown explicitly from the gauge
invariant perturbation theory that, in general, this is not true.  The
RHS of (\ref{eq:SPerTP}) vanishes in the super-Hubble scales (i.e $k
\to 0$) where the perturbations can be treated classical. Hence, as
expected, the trans-Planckian effects are negligible. Secondly, it is
clear from Eq. (\ref{eq:SPerTP}) that the terms in the RHS will
dominate during the trans-Planckian regime and can have interesting
consequences on the primordial spectrum. Lastly, the perturbations (in
general) are not purely adiabatic, i. e., it contains isocurvature
perturbations. However, these perturbations does not contribute
significantly in the super-Hubble scales. Taking the Fourier
transformation of the non-adiabatic part of the pressure perturbation
($\d p_{\rm nad}$), we have
\beq
{\cal F}(\delta p_{\rm nad}) = - 4 d_1 \frac{k^2}{a^3} Q_k^{(2)} \, .
\eeq
\indent From the above expression, it is straight forward to see that, in the
super-horizon scales, the entropic perturbations vanish. Following
Refs. \cite{Wands-Mali:2000,Malik-Wand:2002}, we can assume that, on
large scales, the total curvature perturbation $\zeta$ is
conserved. As mentioned earlier, in the FRW background {\it only} the
canonical scalar field contributes to the stress-tensor.  Following
Ref.\cite{Malik-Wand:2004}, it is possible to show that, on large
scales, {\it only} the curvature perturbation associated to $\d\phi$
contributes to the total curvature perturbation. Hence, it is
sufficient to calculate the power-spectrum associated to the
scalar-field perturbation ($\d\vp$). This will be discussed in 
Sec. (\ref{sec:Pow-Spe})

%%%%%%%%%%%%%%%%%%%%%%%%%%%%%%%%%%%%%%%%%%%%%%%%%%%%%%%%%%%%%%%%%%%%%%%%%%
%%%%%%%%%%%%%%%%%%%%%%%%%  NEW SECTION %%%%%%%%%%%%%%%%%%%%%%%%%%%%%%%%%%%
%%%%%%%%%%%%%%%%%%%%%%%%%%%%%%%%%%%%%%%%%%%%%%%%%%%%%%%%%%%%%%%%%%%%%%%%%%

\section{Classical Analysis}
\label{sec:clas-ana}

In this section, we combine
Eqs. (\ref{eq:MDR-Phi},\ref{eq:per-vpTP},\ref{eq:per-uTP}) to obtain a
single differential equation of $\Phi$. We show that the resultant
differential equation of $\Phi$ is different from that of the standard
canonical scalar field driven inflation. More importantly, the
differential equation of $\Phi$ in our model is fourth order while in
the standard canonical scalar field it is second order. We obtain the
solutions of $\Phi$ in three regimes --- trans-Planckian (I), linear
(II) and super-Hubble (III) --- for the power-law inflation. In the
following section, we obtain the power-spectrum of the scalar
perturbations, in a particular limit, for the power-law inflation.

\subsection{The Power law inflation.}

Let us decompose the fields in their Fourier modes:
\begin{eqnarray}
\Phi(\eta,\vec{x}) &=& \Phi_k(\eta) e^{i \vec{k}.\vec{x}} \, , \,
\delta \varphi(\eta,\vec{x}) = \delta \varphi_k(\eta) e^{i
\vec{k}.\vec{x}} \nonumber\\
%%%
\delta \xi(\eta,\vec{x}) &=& \xi_k(\eta) e^{i \vec{k}.\vec{x}} \, . 
\end{eqnarray}
We have dropped the superscript indicating that the quantities are
gauge invariant.  Combining the equations, we end up with a fourth
order differential equation in the Bardeen potential. To keep the
presentation light, the derivation of the equation is given in
Appendix C.  This equation, for the power-law inflation
(\ref{eq:polaw},\ref{eq:PLaw-para}), reads
{\small
\begin{eqnarray}
 \Phi^{(4)}_k &+& \frac{\Gamma_1}{\eta} \Phi^{(3)}_k \\
%%%%
&+& \left[ \frac{\Gamma_2}{\eta^2} + \left( 1 + \frac{\Gamma_3}{(-\eta)^{(5+3\beta)}} 
\right) k^2 + \frac{\Gamma_4}{(-\eta)^{2(1+\beta)}} k^4 \right] \Phi^{(2)}_k \nn\\
%%%%
&+& \left[ \frac{\Gamma_5}{\eta^3} + \left( \frac{\Gamma_6}{(-\eta)^{3(2+\beta)}}+ 
\frac{\Gamma_7}{\eta} \right) k^2 +
\frac{\Gamma_8}{(-\eta)^{3+2 \beta}} k^4 \right] \Phi^{'}_k \nn \\
%%%%%
&+& \left[ \frac{\Gamma_{9}}{\eta^4} + \frac{\Gamma_{10}}{\eta^2} k^2 +
\left( \frac{\Gamma_{11}}{(-\eta)^{2 \beta + 4}} + 
 \frac{\Gamma_{12}}{(-\eta)^{5 + 3 \beta }} \right) k^4 \right] \Phi_k = 0 \, .
\nn
\end{eqnarray}
}

\noindent 
The constants $\Gamma_i$ depend on the background and the fundamental
constants in the following way
\begin{eqnarray}
\Gamma_1 &=& 4 (2 + \beta) \, , \quad
\Gamma_2 = 12 + 13 \beta + 3 \beta^2 - 6 \sigma_o^2 \, , \nn\\
\Gamma_3 &=& - \frac{3}{2} \frac{(-1)^{3 \beta} \eta_o^{3+3 \beta} 
\sigma_o^2 }{a_o^3} \,  \frac{b_{11}}{d_1} M_{pl}^2  \, , \nn\\
\Gamma_4 &=& 2 \, \frac{(-1)^{2 \beta} \eta_o^{2+2 \beta} }{a_o^2} b_{11} \, , 
\quad
\Gamma_5 = - 18 (1+\beta) \sigma_o^2  \, , \nn\\
\Gamma_6 &=&  3 \frac{(-1)^{3 \beta} \eta_o^{3+3 \beta} 
\sigma_o^2 }{a_o^3} \,  \frac{b_{11}}{d_1} M_{pl}^2 \, , \quad
\Gamma_7 = 6 + 4 \beta \, , \nn\\
\Gamma_8 &=& - 4 (2 + \beta) \, \frac{(-1)^{2 \beta} \eta_o^{2+2 \beta} }{a_o^2} 
b_{11} \, , \nn\\
\Gamma_9 &=& 6 a_o^2 v_o \eta_o^2  \frac{q^2(-24+q^2)}{(-6+q^2)^2} M_{pl}^2
\, , \quad \Gamma_{10} = 4 + 7 \beta + 3 \beta^2 \, , \nn\\
\Gamma_{11} &=& 2(1+\beta) (2 + \beta) \, \frac{(-1)^{2 \beta} \eta_o^{2+2 \beta} }{a_o^2} 
\, b_{11} \, , \nn\\
\Gamma_{12} &=&  \frac{3}{2} \frac{(-1)^{3 \beta} \eta_o^{3+3 \beta} 
\sigma_o^2 }{a_o^3} \,  \frac{b_{11}}{d_1} (2 d_1 - M_{pl}^2)  \quad . 
\end{eqnarray}

\subsection{The zeroth order approximation}

We can find approximate solutions for the power law inflation in the
following way. Introducing the quantity $\epsilon$ by
\be
 \beta = -2 - \epsilon  \, ,
\ee
($\epsilon$ vanishes on the de Sitter space) we can make Taylor
expansions of the coefficients $\Gamma_i$ and postulate the same for
the Bardeen potential
\be
  \Phi_k(\eta) = \sum_{m=0} \epsilon^m \Phi_{k,m}(\eta) \quad .
\ee
The outcome is the following. Each component $\Phi_{k,m}(\eta)$ obeys
a differential equation which is inhomogeneous, the source term
depending on the preceding components $\Phi_{k,0}(\eta), \cdots,
\Phi_{k,m-1}(\eta)$.

As discussed in Ref. \cite{Martin-Bran:2003}, we have to be careful in
the limit of $\epsilon \rightarrow 0$ in the sense that it does not
give the perturbation corresponding to the de Sitter space. This is
related to the fact that on this background the inflaton field is
constant so that the quantities $X_i,Y_i,Z_i$ are undetermined. To
work out what happens for the de Sitter space, one would have to
consider the earlier equations where divisions by the derivatives of
the inflaton was not performed yet.

The zeroth order contribution is a solution of the equation
\begin{eqnarray}
\label{eq:zeroth-ord} 
\Phi^{(4)}_{k,0}(\eta) & + &
  \left( k^2 - \frac{2}{\eta^2} + \gamma^2 k^4 \eta^2  \right) 
  \Phi^{(2)}_{k,0}(\eta)
  - 2 \frac{k^2}{\eta} \Phi^{'}_{k,0}(\eta) \nn\\
 &+& 2 \frac{k^2}{\eta^2} \Phi^{'}_{k,0}(\eta) = 0 \quad .
\end{eqnarray}
We now introduce the dimensionless variable $x$ defined by $x= k \eta$
and the function $f(x)$ given by
\begin{eqnarray}
  \Phi_{k,0}(\eta) = f(x) \quad .
\end{eqnarray}
The fourth order equation takes the form
{\small
\be
f^{(4)}(x) + \left( 1 - \frac{2}{x^2} + \gamma^2 x^2 \right)
f''(x) - \frac{2}{x} f'(x) + \frac{2}{x^2} f(x) = 0 \, , 
\label{eq:fxde}
\ee
}
where,
\be
\gamma = \sqrt{2 b_{11} v_0} M_{pl} \, , \nn
\ee 
As mentioned earlier, we obtain approximate solutions to the above
differential equation in three different regions.

\subsubsection{The first region} 
  
In this region, the term $\gamma^2 x^2 $ dominates. In other words,
the trans-Planckian effects are dominant and we are dealing with large
values of $x$. Using the fact that $f(x)=x$ is a solution of the full
equation, one introduces the function $h(x)$ by the relation 
\be f(x)=
x \int^x h(\zeta) d\zeta 
\ee 
and obtains the third order equation 
\be
2 (-1+\gamma^2 x^2) h(x) + \gamma^2 x^3 h{'}(x) + 4 h{''}(x) + x
h^{'''}(x) = 0 
\ee 
We can get rid of the second derivative by the change of function
\begin{eqnarray} h(x) &=&
\frac{1}{x^{4/3}} R(x) \quad ; \end{eqnarray} 
using the fact that we are in the region given by large values of $x$,
one ends up with the differential equation
\begin{eqnarray} R^{'''}(x) + \gamma^2 x^2
R^{'}(x) + \frac{2}{3} \gamma^2 x R(x) = 0 \end{eqnarray} 
whose solution is a combination of generalized hypergeometric
functions multiplied by polynomials:
\begin{eqnarray} R(x) &=& C^{'}_1 F_{pq}
\left[ \left\{ \frac{1}{6} \right\} , \left\{ \frac{1}{2} ,
\frac{3}{4} \right\} , - \frac{1}{16} \gamma^2 x^4 \right] \nn\\ &+&
C^{'}_2 \sqrt{\gamma} x F_{pq} \left[ \left\{ \frac{5}{12} \right\} ,
\left\{ \frac{3}{4} , \frac{5}{4} \right\} , - \frac{1}{16} \gamma^2
x^4 \right] \nn\\ &+& C^{'}_3 \gamma x^2 F_{pq} \left[ \left\{
\frac{2}{3} \right\} , \left\{ \frac{5}{4} , \frac{3}{2} \right\} , -
\frac{1}{16} \gamma^2 x^4 \right] \, , 
\end{eqnarray} 
where $C'_i$'s are constants to be determined and $F_{pq}$ are the
generalized Hypergeometric functions. Thus, the solution to the differential
equation (\ref{eq:fxde}) is 
%, one finds the scalar perturbation to be given in terms of
%more generalized hypergeometric functions:
%
\begin{eqnarray} f(x) &=& C_1(k) \, x \nn\\ &+& C_2(k) \, x^{2/3}
F_{pq} \left[ \left\{ - \frac{1}{12} , \frac{1}{6} \right\} , \left\{
\frac{1}{2} , \frac{3}{4} , \frac{11}{12} \right\} , - \frac{1}{16}
\gamma^2 x^4 \right] \nn\\ &+& C_3(k) \, x^{5/3} F_{pq} \left[ \left\{
\frac{1}{6} , \frac{5}{12} \right\} , \left\{ \frac{3}{4} ,
\frac{7}{6} , \frac{5}{4} \right\} , - \frac{1}{16} \gamma^2 x^4
\right] \nn\\ &+& C_4(k) \, x^{8/3} F_{pq} \left[ \left\{ \frac{5}{12}
, \frac{2}{3} \right\} , \left\{ \frac{5}{4} , \frac{17}{12} ,
\frac{3}{2} \right\} , - \frac{1}{16} \gamma^2 x^4 \right] \, \nn \\
\end{eqnarray}
where $C_i$'s are related to $C'_i$'s. These generalized
hypergeometric functions have a few properties which are worth
mentioning. First, they are highly oscillating. For example, the
function
\begin{eqnarray}
 F_{pq} \left[ \left\{ - \frac{1}{12} , \frac{1}{6} \right\} ,
   \left\{ \frac{1}{2} , \frac{3}{4} , \frac{11}{12} \right\} , - \frac{1}{16} \gamma^2 x^4 
   \right] 
\end{eqnarray}
goes from $7.7 \, 10^{32}$ to $1.4 \, 10^{194}$ when $x$ goes from $x=50$ to $x=100$,
fixing $\gamma=1/10$ for illustrative purposes.

Let us now say a few words about these generalized hypergeometric
functions; this will help us to quantify their oscillatory
behavior. They are special cases of Meijer functions which can be
defined by integrals on the complex plane \cite{Mathai-Saxe:1973-bk}:
\be
 G^{m,n}_{p,q} \left( z \vert\begin{array}{cccc}
                         a_1 & a_2 & ... & a_p \\
			 b_1 & b_2 & ... & b_q
                      \end{array}   \right)  =
	\frac{1}{2 \pi i}     \int_C  \chi(s) z^{-s} ds
\ee
where
\be
 \chi(s) =  \frac{\Pi_{j=1}^{m} \Gamma(b_j+s) \Pi_{j=1}^{n} \Gamma(1-a_j-s)}
 {\Pi_{j=m+1}^{q} \Gamma(1-b_j-s) \Pi_{j=n+1}^{p} \Gamma(a_j+s)}
\ee
and three possibilities are allowed for the contour $C$, according to some
conditions on the parameters $a_i,b_j,m,n,p,q$ \cite{Mathai-Saxe:1973-bk}.
Our solutions correspond to $m=n=0$.

The asymptotic behavior which is relevant here is the following.
For large values with $- (\nu^\star +1) \pi < arg z < 0$, the dominant 
part is roughly given by
\be
 G^{m,n}_{p,q} \left( z \vert\begin{array}{cccc}
                         a_1 & a_2 & ... & a_p \\
			 b_1 & b_2 & ... & b_q
                      \end{array}   \right)
\sim      H_{p q}(z e^{i \pi \mu^\star} ) \quad .
\ee
where
\begin{eqnarray}
  \mu^\star &=& q - m - n  \, , \quad
  \nu^\star = - p + m + n  \, , \nn\\
  H_{pq}(z) & = & \exp{\left( (p-q)  z^{\frac{1}{q-p}} \right) }  
  z^{\rho^\star}  \, , {\rm and} \nn\\
  \rho^\star &=&  \frac{1}{q-p} 
  \left( \sum_{j=1}^q b_j - \sum_{j=1}^{p} a_j + \frac{p-q+1}{2} \right)
\end{eqnarray}

In our case, one has to make the replacements
\be
  z = - \frac{1}{16} \gamma^2 k^4 \eta^4 \, , \quad q=3 , \, ,
  p = 2 \quad  
\ee
so that
\begin{eqnarray}
\Phi_{k,0}(\eta) &=&  
C_0(k) \, k \eta + \sum_{i=1}^3  C_i(k) \, (\k \eta)^{\sigma_i} \\
&\times &  \left( - \frac{1}{16} \gamma^2 k^4 \eta^4 \right)^{\rho^\star_i} 
\exp{ \left(  \frac{1}{16} \gamma^2 k^4 \eta^4   \exp{(i \pi \mu^\star_i)} 
  \right) }  \, , \nn 
\end{eqnarray}
where $\sigma_1 = 2/3, \sigma_2 = 5/3, \sigma_3 = 8/3$.  Fomr the
above expressiom, it is easy to see that the solution in the the
Bardeen potential $\Phi$ is oscillating in this region.
%The oscillations are easily seen from this formula. 
The choice of the constants $C_i(k)$ correspond to different choices
of initial conditions and thus, in principle, to different choices of
vacua.  We will come back to this later.

\subsubsection{The second region} 

In the intermediary region, $1$ dominates over $\gamma^2 x^2$. The
solution in this region is
\begin{eqnarray} 
f(x) &=& D_1(k) \, x + D_2(k) \, x^2 \nn\\
%%%
&+& D_{3}(k) \, \l[ e^{-i x} (-1+ i x ) - x^2 Ei(-ix) \r] \nn\\
%%%
&+& D_{4}(k) \, \l[ e^{i x} (i - x ) + i x^2 Ei(ix) \r] \,
\end{eqnarray} 
where $Ei(x)$ refers to the exponential integral. Using the asymptotic
behavior of the exponential integral
(cf. Ref. \cite{Abramowitz-Steg:1964-bk}, p. 231), we get
\begin{eqnarray} 
f(x) &=& D_1(k) \, x + D_2(k) \, x^2 \nn\\
%%%
&+& D_{3}(k) \, \l[- e^{-i x} + 2 x \sin(x) \r] \nn\\
%%%
&+& D_{4}(k) \, \l[ i \, e^{i x} - 2 x \sin(x) \r] \,
\end{eqnarray} 
As we can see, the Bardeen potential is a sum of plane-waves. 

\subsubsection{The third region} 

When the term $-2/x^2$ dominates in the coefficient of the second
derivative, the solution can be found and is given by
%by the same reasoning than in
%the first region:
%
\begin{eqnarray}
f(x) &=& G_1(k) + G_4(k) \,  x + G_3(k) \, x^4 + G_2(k) \, x \ln{x} \, .~~
\end{eqnarray}
From the above expression, we see that in the super-Hubble scales the
scalar perturbations has a constant term which is identical to the
canonical scalar field inflation.  

To finish this section, let us remark that in the non trans-Planckian
region, i.e  when 
\[ 1 - \frac{2}{x^2} >> \gamma^2 x^2  \, , \]  
the solution to the differential equation (\ref{eq:fxde}) can be obtained 
and is given by
\begin{eqnarray}
   f(x) &=& H_1(k) \, x + \frac{1}{2}\,  H_2(k) \, (2- 2 x+x^2) \nonumber\\
 &+& \frac{1}{2} \, H_3(k) \, x \left(- \frac{e^{-i x}}{x} - i Ei(- i x) \right) \nonumber\\ 
 &+& \frac{1}{2} \, H_4(k) \,  x \left( \frac{e^{i x}}{x} - 
 i Ei(i x) \right) \quad ;
\end{eqnarray}
this approximation covers the $II$ and $III$ region simultaneously.

\subsection{The first order approximation}

The first order contribution obeys the equation
\begin{eqnarray}
& & \Phi_{k,1}^{(4)}(\eta) 
+ \left( k^2 - \frac{2}{\eta^2} + \gamma^2 k^4 \eta^2  \right) 
\Phi_{k,1}^{(2)}(\eta) - 2 \frac{k^2}{\eta} \Phi_{k,1}^{'}(\eta) \nn\\
%%%%%%
& & \qquad \qquad +~2 \frac{k^2}{\eta^2} \Phi_{k,1}(\eta) = S_k(\eta) \, ,
\end{eqnarray}
where 
\begin{eqnarray}
S_k(\eta) &=& - \frac{4}{\eta}  \Phi_{k,0}^{(3)}(\eta) 
+  \left[ - \frac{5}{\eta^2} + 
\frac{b_{_{11}} M_{pl}^5 v_o^{3/2}}{d_1} k^2 \eta \right. \nn\\
%%%%%
&- & \left.  \frac{10}{3} b_{_{11}} M_{pl}^2 v_o k^4 \eta^2  
 + 4 b_{11} M_{pl}^2 v_o \eta^2  \log{ \left(\frac{\eta}{\eta_o}
 \right)}\right] \Phi_{k,0}^{(2)}(\eta) \nn\\
%%%%
&+ &  \left[\frac{12}{\eta^3} - 4 \frac{k^2}{\eta} - 
4 b_{11} M_{pl}^2 v_o k^4\eta  \right] \Phi_{k,0}^{'}(\eta)  \nn\\
%%%%
&+& \left[ - \frac{24}{\eta^4} + \frac{5 k^2}{\eta^2} 
+ 2 b_{11} M_{pl}^2 v_o  \right. \nn\\
%%%%
& & \left. + \frac{b_{11} M_{pl}^3 v_o^{3/2} (- 2 d_1 +
M_{pl}^2 ) }{d_1} \eta  \right] \Phi_{k,0}{(\eta)}  \, .\nn
\end{eqnarray}
This equation is exactly the one obeyed by the zeroth order
contribution, except for the source term which is known since we
obtained the approximations of the zeroth order in the three
regions. Let us specialize to one of the regions and call
$Y_1(\eta),Y_2(\eta),Y_3(\eta), Y_4(\eta)$ the four different
solutions of the homogeneous equation given in Eq~(\ref{eq:zeroth-ord}):

The equation being linear and knowing the complete solution of the
homogeneous equation ($ \Phi=\sum_{a=1}^4 L_a Y_a $ with $L_a$
constants) solution, we can solve it using the method of the variation
of the constants.  One can show that this can be achieved by the following 
system of equations
\begin{eqnarray} 
\sum_{a=1}^4    L^{'}_a Y_a = 0 &,& 
\sum_{a=1}^4    L^{'}_a Y^{'}_a = 0 \, , \nn \\
\sum_{a=1}^4       L^{'}_a Y^{(2)}_a = 0 
&,&  \sum_{a=1}^4   L^{'}_a Y^{(3)}_a = S(\eta) \, .
\end{eqnarray}
Let us concentrate on the second and third region for example
(the non trans-Planckian zone). One has
\begin{eqnarray}
\Phi_{k,1}(\zeta) &=& k \zeta \int_0^{\zeta} d\eta 
\left[ i \frac{(2 i+2  k \eta - i k^2 \eta^2)}{2 k^4 \eta} 
Ei(i k \eta) \right. \nn\\
%%%%%
&+& \left. \frac{(-2 i + 2 k \eta + i k^2 \eta^2)}{2 k^4 \eta} Ei(-i k
\eta) \right] S_k(\eta)  \nn\\
%%%%%
&+&\left[ 1- k \zeta +\frac{1}{2} k^2 \zeta^2 \right] \int_0^\zeta
d\eta \, 2 \, \frac{1}{k^4 \eta} S_k(\eta)  \nn\\
%%%%
&+& \left[ - \frac{1}{2} i e^{-i k \zeta} + \frac{1}{2} x Ei(-i k
\zeta) \right] \nn\\
%%%%%
& & \times \int_0^\zeta d\eta \, i \, \frac{(-2+2 i k \eta + k^2
\eta^2)}{k^4 \eta} e^{i k \eta} S_k(\eta)  \nn\\
%%%%%
&+& \left( \frac{1}{4} i e^{i k \zeta} - \frac{1}{4} i k \zeta Ei(i k
\zeta) \right) \\
%%%%%
& & \times \int_0^\zeta d\eta \, 2 \frac{(-2-2 i k \eta + k^2
  \eta^2)}{k^4 \eta} e^{-i k \eta} S_k(\eta) \, .  \nn
\end{eqnarray}
A similar treatment can be applied to the trans-Planckian region but
the formulas are too lengthy and will not be recorded here.

Using the analysis discussed in this section, the power spectrum of
the perturbations can be obtained upto a $k$ dependent constant
factor. In order to obtain the exact power spectrum, we need to
quantize the theory and fix the initial state of the field
\cite{Mukhanov-Feld:1992}. In the following section, we obtain exact
power spectrum of the perturbations in a particular limit.

%%%%%%%%%%%%%%%%%%%%%%%%%%%%%%%%%%%%%%%%%%%%%%%%%%%%%%%%%%%%%%%%%%%%%%%%%%
%%%%%%%%%%%%%%%%%%%%%%%%%  NEW SECTION %%%%%%%%%%%%%%%%%%%%%%%%%%%%%%%%%%%
%%%%%%%%%%%%%%%%%%%%%%%%%%%%%%%%%%%%%%%%%%%%%%%%%%%%%%%%%%%%%%%%%%%%%%%%%%

\section{Power-spectrum of the perturbations -- Quantum Analysis}
\label{sec:Pow-Spe}

In this section, we calculate the power-spectrum corresponding to
$\mu_{_S}$ during the power-law inflation using the following
approach: (i) We assume that the quantum field $\mu_{_S}$ is coupled
to an external, classical source field ${\cal S}_k(\eta)$ which is
determined by solving the coupled differential equations
(\ref{eq:MDR-Phi}, \ref{eq:per-uTP}). (ii) We solve the equation of
motion of $\mu_{_S}$ in three regions -- Trans-Planckian (I), linear
(II) and super-Hubble (III) -- separately \cite{Martin-Bran:2000}. We
further assume that ${\cal S}_k(\eta)$ will contribute significantly
in the trans-Planckian region while it can be neglected in the linear
and super-Hubble region. (iii) The power-spectrum at the super-Hubble
scales is determined by performing the matching of the modes and its
derivatives at the times of transition between regions I and II
[$(-\eta_{\rm Pl})^{1 + \b} \equiv (\omega k^2)^{-1/2}$] and regions II and III
[$\eta_{_H} \equiv (1 + \b)/k$]. We assume that the quantum field
$\mu_{_S}$ is in a minimum energy state at $\eta = \eta_i$
\cite{Brown-Dutt:1978}.

Region (I) corresponds to the limit where the non-linearities of the
dispersion relation play a dominant role, i. e. $2b_{11} [k/a(\eta)]^2
\gg 1$ and $k \eta \gg 1 $.  Region (II) corresponds to the limit
where the non-linearities of the modes are negligible i. e. $\omega
\simeq k$ and $k \eta \gg 1 $.  Region (III) corresponds to the limit
where $k \eta \ll 1$. In the three regions, the equation of motion of
$\mu_{_S}$ (\ref{eq:SPerTP}) reduces to:
\begin{subequations}
\label{eq:SPerTP-Reg}
\br
\label{eq:SPerTP-Reg1}
{\mu_{_S}^{(I)}}^{''} + \omega^2(\eta) \mu_{_S}^{(I)} &\simeq& 
- \frac{2 d_1}{\MPl} k^2 {\cal S}_{k}(\eta) \, , \\
\label{eq:SPerTP-Reg2}
{\mu_{_S}^{(II)}}^{''} +  k^2 \mu_{_S}^{(II)} &\simeq&  0 \, , \\
\label{eq:SPerTP-Reg3}
{\mu_{_S}^{(III)}}^{''} -   
\frac{\beta (\beta + 1)}{\eta^2} \mu_{_S}^{(III)} &\simeq&  0  \, ,
\er
\end{subequations}
where 
\beq
\omega(\eta) =  \omega_0 \, \frac{k^2}{(-\eta)^{(1 + \beta)}} \, ; \,
\omega_0 = (2 b_{_{11}})^{1/2} (-\eta_0)^{(1 + \beta)} \, ,
\eeq
and ${\cal S}_k$ is given by Eq. (\ref{eq:sou-fin}). The general
solution to the differential equation (\ref{eq:SPerTP-Reg1}) is given
by
\br
\mu^{\rm(I)}_{_S}(\eta ) & = & A_1(k)\, (-\eta)^{1/2} 
H_{\nu}^{(1)}[\alpha(\eta)] \nn \\
%%%%
&+& A_2(k) \, (-\eta)^{1/2} H_{\nu}^{(2)}[\alpha(\eta)] + \mu_{_P}(\eta)
\label{eq:Sol-Reg1}
\er
where $\mu_{_P}(\eta)$ is the particular solution to the inhomogeneous
part of the differential equation and is given by
(cf. Ref. \cite{Morse-Fesh:1953-bk}, p. 529)
\begin{widetext}
{\small \beq
\label{eq:par-sol}
\mu_{_P}(\eta) = \frac{i \pi}{2} \frac{d_1 k^2}{\beta \MPl}  
(-\eta)^{1/2} \l[ H_{\nu}^{(1)}[\alpha(\eta)] 
\int_{\eta_l}^{\eta} (-s)^{1/2} H_{\nu}^{(2)}[\alpha(s)] {\cal S}_k(s) \, ds 
+  H_{\nu}^{(2)}[\alpha(\eta)] 
\int_{\eta_l}^{\eta} (-s)^{1/2} H_{\nu}^{(1)}[\alpha(s)] {\cal S}_k(s) \, ds 
\r]
\eeq
}
\end{widetext}
\beq
\label{eq:def-nu-a}
\! {\rm and} \quad
\nu = - \frac{1}{2 \beta};~~~
\alpha(\eta) = \alpha_0 (-\eta)^{-\b};~~~
\alpha_0 = \frac{\omega_0 k^2}{-\beta} 
\, ,
\eeq
$\eta_l$ ($<\eta_i$) is the epoch in which the integrals in
(\ref{eq:par-sol}) vanish. The quantities $H_{\nu}^{(1)}$ and
$H_{\nu}^{(2)}$ in the above solution are the Hankel functions of the
first and the second kind (of order $\nu$), respectively, and the
$k$-dependent constants $A_1(k)$ and $A_2(k)$ are to be fixed by the
initial conditions for the modes at $\eta _{\rm i}$. Unlike the
canonical scalar field inflation, where one assumes that the field is
in a Bunch-Davies vacuum at $\eta_i$, it is not possible to assume
such an initial condition due to the non-linearities of the modes. As
mentioned earlier, we assume that the field is in the minimum energy
vacuum state at $\eta_i$, i. e.,
\begin{equation}
\label{eq:Ini-Sta}
\mu_{_S}(\eta _{\rm i}) = \frac{1}{\sqrt{2 \omega(\eta_{\rm i})}}~;~
\mu_{_S}'(\eta _{\rm i})= \pm i \sqrt{\frac{\omega(\eta_{\rm i})}{2}}.
\end{equation}
We thus get
\begin{subequations}
\begin{eqnarray}
\label{Ai1sol}
A_1(k) &=& \frac{i \pi  \alpha(\eta_i)}{4} (-\eta _{\rm i})^{-1/2}
{\tilde \mu_{_S}(\eta_i)} H_{\nu - 1}^{(2)}[\a(\eta_i)] \nn \\
%%%%
& \times& \biggr[1 + \frac{(-\eta _{\rm i})^{\b + 1}}{(-\a_0 \b)}
\frac{{\tilde \mu_{_S}}'(\eta _{\rm i})}{{\tilde \mu_{_S}}(\eta _{\rm i})}
\frac{H_{\nu}^{(2)}[\a(\eta_i)]}{H_{\nu - 1}^{(2)}[\a(\eta_i)]}\biggr], \\
\label{Ai2sol}
A_2(k) &=& - \frac{i \pi  \alpha(\eta_i)}{4} (-\eta _{\rm i})^{-1/2}
{\tilde \mu_{_S}(\eta_i)} H_{\nu - 1}^{(1)}[\a(\eta_i)] \nn \\
%%%%
& \times& \biggr[1 + \frac{(-\eta _{\rm i})^{\b + 1}}{(-\a_0 \b)}
\frac{{\tilde \mu_{_S}}'(\eta _{\rm i})}{{\tilde \mu_{_S}}(\eta _{\rm i})}
\frac{H_{\nu}^{(1)}[\a(\eta_i)]}{H_{\nu - 1}^{(1)}[\a(\eta_i)]}\biggr] \, ,
\end{eqnarray}
\end{subequations}
where ${\tilde \mu_{_S}}(\eta) = \mu_{_S}^{(I)}(\eta) -
\mu_{_P}(\eta)$.  From the above expressions it is evident that the
particular solution to the inhomogeneous differential equation
effectively changes the initial state of the field. Hence, ${\tilde
\mu_{_S}}(\eta_{\rm i})$ can be treated as the new effective initial
state of the field. It is worth mentioning that the particular
solution has not been fixed and is arbitrary.

In Region II, the solution to the differential equation
(\ref{eq:SPerTP-Reg2}) is (Minkowski) plane waves, i. e.,
\begin{equation}
\label{eq:Sol-Reg2}
\mu_{_S}^{\rm (II)}(\eta ) \, = \, B_1(k) \exp[-i k \eta] + 
B_2(k) \exp[i k \eta] \, ,
\end{equation}
where $B_1(k), B_2(k)$ are $k-$dependent constants and are obtained by
the junction conditions of the mode functions $\mu_{_S}^{\rm (I)},
\mu_{_S}^{\rm (II)}$ and their derivatives at $\eta = \eta_{\rm Pl}$. This
gives,
\begin{widetext}
{\small
\begin{subequations}
\begin{eqnarray}
\label{B1}
\!\!\!\!\!\!\!\!\!\!\!\!\!\!
\frac{\exp(-i k \eta _{\rm Pl})}{(-\eta_{\rm Pl})^{1/2}} B_1 &=& 
\frac{A_1}{2}  H_{\nu}^{(1)}[\a(\eta_{\rm Pl})] 
\l[1 + \frac{i \b \a_0}{k (-\eta_{\rm Pl})^{\beta + 1}} 
\frac{H_{\nu - 1}^{(1)}[\a(\eta_{\rm Pl})]}{H_{\nu}^{(1)}[\a(\eta_{\rm Pl})]}
\r]  
+ \frac{A_2}{2} H_{\nu}^{(2)}[\a(\eta_{\rm Pl})] 
\l[1 + \frac{i \b \a_0}{k (-\eta_{\rm Pl})^{\beta + 1}} 
\frac{H_{\nu - 1}^{(2)}[\a(\eta_{\rm Pl})]}{H_{\nu}^{(2)}[\a(\eta_{\rm Pl})]}
\r] \, , ~~\\
%%%%%
\label{B2}
\!\!\!\!\!\!\!\!\!\!\!\!\!\! 
\frac{\exp(i k \eta _{\rm Pl})}{(-\eta_{\rm Pl})^{1/2}} B_2 &=& 
\frac{A_1}{2}  H_{\nu}^{(1)}[\a(\eta_{\rm Pl})] 
\l[1 - \frac{i \b \a_0}{k (-\eta_{\rm Pl})^{\beta + 1}} 
\frac{H_{\nu - 1}^{(1)}[\a(\eta_{\rm Pl})]}{H_{\nu}^{(1)}[\a(\eta_{\rm Pl})]}
\r]  
+ \frac{A_2}{2} H_{\nu}^{(2)}[\a(\eta_{\rm Pl})] 
\l[1 - \frac{i \b \a_0}{k (-\eta_{\rm Pl})^{\beta + 1}} 
\frac{H_{\nu - 1}^{(2)}[\a(\eta_{\rm Pl})]}{H_{\nu}^{(2)}[\a(\eta_{\rm Pl})]}
\r] \, . ~~
\end{eqnarray}
\end{subequations}
}
\end{widetext}
In region III, the solution is 
\begin{equation}
\mu _{\rm (III)}(\eta ) \, = \, C(k) \, a(\eta ) \, ,
\end{equation} 
where $C(k)$ is constant (not to be confused with the constants used
in the previous section) whose modulus square gives the power spectrum
of the density perturbations and is determined by performing the
matching of the modes $\mu_{_S}^{\rm (II)}, \mu_{_S}^{\rm (III)} $ at
$\eta_{_H} \equiv (1 + \b)/k$. We thus get,
{\small
\beq
\!\!\!\! C(k) =  \l[\frac{\eta_{_0} k}{1 + \b}\r]^{1 + \b} \!\!\!
\l[B_1(k) \exp(-i k \eta_{_H}) + B_2(k) \exp(i k \eta_{_H})\r] \, .
\eeq
}
\noindent The spectrum of the perturbations (\ref{eq:pS0}) reduce to
\beq 
\left[k^3\; {\cal P}_{S}(k)\right]
=\left(\frac{1}{4 \pi^2 \MPl} \frac{(\b + 1)}{\b}\r) \, k^3
\l|C(k)\r|^2\, .
\label{eq:pS}
\eeq 
We are interested in the leading order behavior of the primordial
power-spectrum and the possible modifications to the primordial
spectrum due to the trans-Planckian effects. In order to do that, we
need to obtain the leading order behavior of the constants $A_1, A_2,
B_1$ and $B_2$.  Using the fact that $k \eta \gg 1$ and the asymptotic
behavior of the Hankel functions,
viz. (cf.~Ref.~\cite{Abramowitz-Steg:1964-bk}, p.~364)
\br
\lim_{z \to \infty} H^{(1/2)}_{\nu}(z)\longrightarrow
\l({\frac{2}{\pi z}}\r)^{1/2}\, {\rm e}^{\pm i\l[z-(\pi \nu /2)-(\pi /4)\r]},
\er
we get
\begin{subequations}
\begin{eqnarray}
\label{A1approxJCr}
A_1(k) &\approx & A_0
 {\tilde \mu_{_S}(\eta_i)} \exp(- i x_{\rm i})
\l( 1 \mp {\cal I}\r) \, , \\
\label{A2approxJCr}
A_2(k) &\approx & A_0 {\tilde \mu_{_S}(\eta_i)} \exp(i x_{\rm i})
\l( 1 \pm  {\cal I}\r) \, , 
\end{eqnarray}
\end{subequations}
where 
\br
A_0= \l(\frac{\pi \alpha_0}{8}\r)^{1/2} (-\eta _{\rm i})^{-(\beta + 1)/2}
&;& x_{\rm i} = \a(\eta_i) - \frac{\pi \nu}{2} - \frac{\pi}{4} \nn \\
{\cal I}= \frac{1 - \mu_{_P}'(\eta_i)/\mu_{_S}'(\eta_i)}
{1 - \mu_{_P}(\eta_i)/\mu_{_S}(\eta_i)} & & \, .
\er
Having obtained $A_1, A_2$ in the limit of $k\eta \gg 1$. Our next
step is to evaluate $B_1(k), B_2(k)$ in the same limit. In order to do
that, we need to know the correct matching time $\eta_{\rm
Pl}$. Demanding $\omega^2(\eta_{_{\rm Pl}}) = k^2$ gives $(-\eta_{\rm
Pl})^{1 + \b} = \omega^{-1/2}/k$. We thus get,
\begin{eqnarray}
\label{B1approxJCr}
B_1 &\approx & A_1 \l(\frac{- 2 \b}{\pi k}\r)^{1/2} \exp(i x_{\rm Pl}) \\
\label{B2approxJCr}
B_2 &\approx & A_2 \l(\frac{- 2 \b}{\pi k}\r)^{1/2} \exp(-i x_{\rm Pl})
\er
where $x_{\rm Pl} = k \eta_{\rm Pl} (\b + 1)/\b - \pi \nu/2 - \pi/4$.
Thus, we get,
\br
\label{eq:SPSPec-Gen}
\left[k^3\; {\cal P}_{S}(k)\right] &\simeq& 
C_0 k^{2(\b + 2)} \l |1 - \frac{\mu_{_P}(\eta_i)}{\mu_{_S}(\eta_i)}\r |^2  \\
& \times & \l[1 + 2 \cos(x_{_H}) 
- 2 Im[{\cal I}] \sin(x_{_H})\r] \nn 
\er
where 
\br
C_0 &=& \left(\frac{1}{16 \pi^2 \MPl} \frac{(\b + 1)}{\b}\r) 
\l(\frac{-\eta_0 \eta_i}{1 + \b}\r)^{2 (1 + \b)}~; \nn \\ 
x_{_H} &=& 2(1 + \b - x_{_{Pl}} + x_i) \, 
\er
and we have neglected higher order terms like $|{\cal I}|^2$.  It is
interesting to note that in the limit of ${\cal S}_k(\eta) \to 0$, the
power-spectrum is same as that of the standard power-law inflation
spectrum with small oscillations. In this limit, we recover the result
of Refs. \cite{Martin-Bran:2000,Niemeyer-Pare:2001}. In order to
obtain the exact form of the power-spectrum, we need to evaluate
$\mu_{_P}$ which requires the knowledge of ${\cal S}_k(\eta)$.

In the rest of this section, we evaluate the power-spectrum in a
particular limit ($1/c_1 \to 0$). We, first, obtain the form of ${\cal
S}_k(\eta)$ by solving the system of coupled differential equations
(\ref{eq:MDR-Phi}, \ref{eq:per-uTP}) in two -- sub-Hubble and
super-Hubble -- regimes. As mentioned earlier, the two differential
equations (\ref{eq:MDR-Phi}, \ref{eq:per-uTP}) do not contain higher
order spatial derivatives. Hence, it is sufficient to obtain solutions
in these two regimes. Performing the following transformations
\beq
u = \frac{a(\eta)}{\ov{\vp}'(\eta)} \Phi~;~
\xi^{(gi)} = a^{3/2}(\eta) \, {\tilde \xi} \, ,
\eeq
and taking the Fourier transform, Eqs. (\ref{eq:MDR-Phi},
\ref{eq:per-uTP}), reduce to
\br
\label{eq:CSuk-1}
& & u_{k}'' + \l(c_s^2 k^2 - \frac{\theta''}{\theta} \r) u_k 
= - \frac{2 d_1}{\MPl} \frac{k^2}{\ov{\vp}'} {\l(a^{3/2} {\tilde \xi}_k\r)}^{'} \\
%%%%%
\label{eq:CSxi-1}
& & {\tilde \xi}'' + \l[ - \frac{9}{4} \Hbar^2 + \frac{3}{2} \Hbar' + 
\frac{c_1}{a^2} \frac{\vp'^{2}}{2 \MPl} \frac{k^2}{\MPl} \r] {\tilde \xi} \\
%%%%
&=& a^{1/2} \l[\l(1 + \frac{c_1}{a^2}\frac{k^2}{\MPl} \r) \Phi_k' 
- 2 \Hbar \l(1 - \frac{c_1}{2 a^2}\frac{k^2}{\MPl} \r) \Phi_k\r] \, .\nn 
\er
In the limit of $1/c_1 \to 0$ (i. e. $d_{1}/\MPl \ll b_{_{11}} \MPl$),
the above differential equations can be solved exactly. In this limit,
the above differential equations become:
\br
& & u_{k}'' + \l(k^2 - \frac{\theta''}{\theta} \r) u_k = 0 \nn \\
& & \frac{\vp'^{2}}{2 \MPl} \xi_k^{(gi)} = a (\ov{\vp}' u_k)'  \, .
\er 
For the sub-Hubble scales, during the power-law inflation, we get
\br
u_{k} &=& D_1(k) \exp(- i k \eta) + D_2(k) \exp(i k \eta) \nn \\
%%%%%
\xi_{k}^{(gi)} &=& 
i k \frac{(-\eta)^{(\b + 2)}}{(-\eta_0)^{(\b + 1)}} 
\sqrt{\frac{2 \MPl}{\b (\b + 1)}}  \nn \\
& & \quad \l[D_1(k) \exp(i k \eta) - D_2(k) \exp(-i k \eta)\r]
\, ,
\er
where we have neglected the terms of the order $1/(k \eta)$ and
$D_{1}(k),D_{2}(k)$ are $k$-dependent constants with the dimensions
of length squared ($k^{-2}$). Using the condition that the modes are
outgoing, we set $D_2(k) = 0$. In the super-Hubble scales, we have
\beq
u_{k} \simeq D_{3}(k) \, a(\eta)~;~ \xi_{k}^{(gi)} =  D_{3}(k)
\sqrt{\frac{2 \b}{\b + 1}} a^2(\eta) \, ,
\eeq
where $D_{3}(k)$ is a constant. In the sub-Hubble scales, we have
\beq
\!
{\cal S}_k(\eta) =  4 \, i \,\MPl b_{_{11}} \, D_1(k) k^5 \, 
\l(\frac{-\eta_0}{- \eta}\r)^{\b + 1} \!\!\!\!\! \exp(-i k \eta) \, .
\eeq

Our next task is to obtain $\mu_{_P}$ and the power-spectrum of the
scalar perturbations. From Eq. (\ref{eq:par-sol}) using the asymptotic 
limit of Hankel functions, we get
\begin{multline}
\mu_{_S}(\eta) = \l(\frac{b_{_{11}}}{\beta^2}\r)^{\frac{\b + 1}{4\b}}
\frac{(-\eta_0)^{\frac{1 - \b^2}{2 \b}}}{(-\eta)^{-\frac{\b +1}{2}}} \, 
\MPl k^{1/\b} \\ 
\times \cos\l[\alpha(\eta) + \ln\l(\Gamma\l[\frac{\b -1}{2 \b}, i
\alpha(\eta)\r]\r)\r] \, ,
\end{multline}
where we have set $D_1 \propto 1/k^2$. Substituting the above
expression in Eq. (\ref{eq:SPSPec-Gen}), we get,
\br
\label{eq:SPSPec-Spe}
\left[k^3\; {\cal P}_{S}(k)\right] &=& 
C_0 k^{2(\b + 2)} \l |1 - C_1 \, k^{(1 + 1/\b)}\r |^2  \\
& \times & \l[1 + 2 \cos(x_{_H}) 
- 2 Im[{\cal I}] \sin(x_{_H})\r] \nn \, ,
\er
where $C_1$ depends on $b_{_{11}}$ and parameters of the power-law
inflation. 
\begin{figure}[!htb]
\begin{center}
\epsfxsize 3.00 in
\epsfysize 2.50 in
\epsfbox{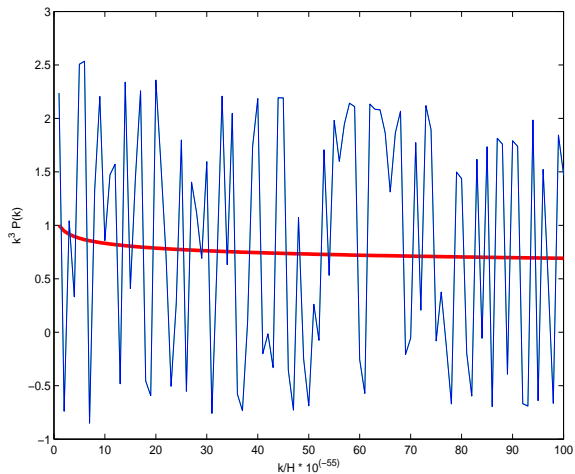}
\caption{Plots of the standard power spectrum (thick curve) and the 
modified power spectrum with appropriate normalization. 
In plotting these spectra we have assumed that $\Hbar=10^{14}\,
{\rm GeV}=10^{52}\, {\rm Mpc}^{-1}$, $(\Hbar/k_{\rm c})= 10^{-4}$ and
$\beta = - 2.04$ where $b_{_{11}} = k_{\rm c}^{-2}$ (cut-off scale) and 
$\eta_{_0} = \Hbar^{-1}$ (inflationary energy scale). The above range 
of $k/\Hbar$ corresponds to $2 < \ell < 100$ where $\ell$ denotes multipoles.}
\label{fig:comp}
\end{center}
\end{figure}

In Fig. (\ref{fig:comp}), we have plotted the standard and the
trans-Planckian inflationary power spectra. As can be seen the
trans-Planckian power spectrum has oscillations. We would like to
caution the readers that the oscillations are small, here we have
magnified the effect for illustrative purposes. We would also like to
point the following: The power-spectrum (\ref{eq:SPSPec-Spe}) we have
obtained becomes significantly different compared to that of
Ref. \cite{Martin-Bran:2000} for very large $k$. For example, for a
particular wave-vector this vanishes.  However such an effect is not 
observable. 

Following points are to be noted regarding the above results: (i) We
have obtained the general power-spectrum (\ref{eq:SPSPec-Gen}) of the
scalar perturbations assuming that the scalar field is in the minimum
energy state and that the contribution of the unit vector field to the
energy density can be neglected. We have shown that the power spectrum
depends on the form of the source term ${\cal S}_k$ which can be
solved analytical in some particular limits. (ii) We have computed the
power-spectrum of perturbations in a particular limit i. e. $(1/c_1
\to 0)$. In this limit, we recover the result of 
Refs. \cite{Martin-Bran:2000,Niemeyer-Pare:2001}. 

%%%%%%%%%%%%%%%%%%%%%%%%%%%%%%%%%%%%%%%%%%%%%%%%%%%%%%%%%%%%%%%%%%%%%%%%%%
%%%%%%%%%%%%%%%%%%%%%%%%%  NEW SECTION %%%%%%%%%%%%%%%%%%%%%%%%%%%%%%%%%%%
%%%%%%%%%%%%%%%%%%%%%%%%%%%%%%%%%%%%%%%%%%%%%%%%%%%%%%%%%%%%%%%%%%%%%%%%%%

\section{Discussion and Conclusion}

In this work, we have computed the gauge-invariant cosmological
perturbation for the single scalar field inflation with the
trans-Planckian effects introduced {\it via} the Jacobson-Corley
dispersion relation. Even though the dispersion relation breaks the
local Lorentz invariance, a covariant formulation of the
corresponding theory can be carried out by introducing a unit
time-like vector field.

Using the covariant Lagrangian, we have obtained the perturbed
stress-tensor for the scalar and tensor perturbations around the FRW
background. We have shown the following: (i) The non-linear effects
introduce corrections to the perturbed energy density while the other
components of the perturbed stress-tensor remains unchanged. Thus,
for the trans-Planckian scenario, we have shown that $\Phi =
\Psi$ and the constraint equation (\ref{eq:hydro-eq2}) remains unchanged. 
(ii) The non-linear terms contributing to the stress-tensor are
proportional to $k^2$ and hence in the super-Hubble scales, as
expected, the contribution to the perturbed energy density can be
ignored. (iii) The spatial higher derivative terms appear {\it only}
in the equation of the motion of the perturbed inflation field
($\d\vp$) while the speed of propagation of the perturbations [in the
equation of motion of the scalar perturbations ($\Phi$)] is different
from that of the standard inflation. (iv) The speed of propagation of
the perturbations ($c_s^2$) is different from that of the canonical
single scalar field inflation. (v) The perturbations are not purely
adiabatic. $\xi$ act as an extra scalar field during inflation and
hence can act as a source. This introduces non-adiabatic (entropic)
perturbations. (vi) Since, the trans-Planckian corrections do not
change the pressure perturbations, the perturbation equations for the
tensor modes do not change. Hence, the tensor perturbation equation
remain unchanged. Recently, Lim \cite{Lim:2004} had show that general
Lorentz violating models (with out taking in-account the higher
derivatives of the scalar field) can modify the pressure perturbations
and hence the tensor perturbation equations. However, in this model,
this is not the case. Since the tensor perturbations remain the same,
the well-know consistency relation between the scalar and tensor ratio
will also be broken in this model \cite{Hui-Kinn:2001}.

We combined Eqs. (\ref{eq:MDR-Phi},\ref{eq:per-vpTP},\ref{eq:per-uTP})
to obtain a single differential equation of $\Phi$. We have shown that
the resultant differential equation of $\Phi$ is different from that
of the standard canonical scalar field driven inflation. More
importantly, the differential equation of $\Phi$ in our model is
fourth order while in the standard canonical scalar field it is second
order. We also obtained the solutions of $\Phi$ in the three regimes
for the power-law inflation.

We have also obtained the equation of motion of the Mukhanov-Sasaki
variable for the perturbed inflaton field with higher derivatives.  In
all the earlier analyzes, the Mukhanov-Sasaki variable ($Q$) was
assumed to satisfy the differential equation Eq. (\ref{eq:SPerTP}) in
which the source term (${\cal S}_k$) was assumed to be zero. More
importantly, we had shown that the source term in
Eq. (\ref{eq:SPerTP}) dominates during the trans-Planckian regime. The
Mukhanov-Sasaki variable of the two fields are strongly coupled and
hence obtaining the solution analytically is possible only in a
particular limit. 

In this work, we calculated the power-spectrum corresponding to the
inflaton field, during the power-law inflation by assuming that (i)
the quantum field $\mu_{_S}$ is coupled to an external, classical
source field ${\cal S}_k(\eta)$ which is determined by solving the
coupled differential equations (\ref{eq:MDR-Phi}, \ref{eq:per-uTP})
(ii) the quantum field is initially in a minimum energy state and
(iii) $d_1/\MPl \ll b_{_{11}} \MPl$. We have shown that in this
particular limit, the power-spectrum is same as that obtained in
Refs. \cite{Martin-Bran:2000,Niemeyer-Pare:2001}.

The work suggests various possible directions for further study:
\begin{itemize}
\item We have obtained the power-spectrum analytically in the 
limit of $d_1/\MPl \ll b_{_{11}} \MPl$. The trans-Planckian
corrections in this limit are small to be observed in the present or
the future CMB experiments. It would be interesting to obtain the
power-spectrum by solving the system of differential equations
numerically and obtain the leading order trans-Planckian corrections
in these models. 
\item As we have mentioned earlier, this model introduces non-adiabatic
perturbations which can lead to the non-Gaussianity in the CMB.
Recently in
Refs. \cite{Martin-Ring:2003a,Martin-Ring:2004a,Martin-Ring:2004b,Easther-Kinn:2004,Easther-Kinn:2005},
trans-Planckian constraints from the CMB was studied in detail. It
would be interesting to do a similar analysis for this scenario. The
non-Gaussian signatures may place stringent and independent
constraints on the parameters $b_{_{11}}, d_{1}$.
\item In this work, we have ignored the solenoidal part 
of the perturbed $u$ field. The solenoidal part contributes to the
vector perturbations. It would be interesting to see whether the
solenoidal part of the perturbed unit-time like vector field can lead
to the growing large-scale vorticity and hence the production of large
scale primordial magnetic field.
\item In this work, we have ignored the back-reaction of 
the field excitations on the perturbed FRW background. There have been
claims in the literature \cite{Tanaka:2000,Starobinsky:2001} that
trans-Planckian modes may effect the evolution of cosmological
fluctuations in the early stages of cosmological inflation in a
non-trivial way. In Ref. \cite{Brandenberger-Mart:2004}, the authors
have discussed in detail the backreaction problem of the
trans-Planckian inflation in a toy model and have shown that the
back-reaction of the trans-Planckian modes may lead to a
renormalization of the cosmological constant driving inflation.. It
would be interesting to perform a similar analysis for this model.
\end{itemize}
We hope to return to study some of these issues in the near future.

%%%%%%%%%%%%%%%%%%%%%%%%%%%%%%%%%%%%%%%%%%%%%%%%%%%%%%%%%%%%%%%%%%%%%%%%%%
%%%%%%%%%%%%%%%%%%%%%%%% ACKNOWLEDGEMENT %%%%%%%%%%%%%%%%%%%%%%%%%%%%%%%%%
%%%%%%%%%%%%%%%%%%%%%%%%%%%%%%%%%%%%%%%%%%%%%%%%%%%%%%%%%%%%%%%%%%%%%%%%%%

\section*{Acknowledgments}
The authors wish to thank J. Martin, L. Sriramkumar for comments on
the earlier version of the paper. The authors also wish to thank
S. Bashinsky, U. Seljak and in particular, N. Bartolo for stimulating
discussions. SS thanks D. Mattingly, B. van Tent for useful email
correspondences.

\appendix
%%%%%%%%%%%%%%%%%%%% NEW APPENDIX %%%%%%%%%%%%%%%%%%%%%%%%%%%%%%%
\section{The Background.}
\label{sec:FRW-bkg}

In this appendix, we give key steps in obtaining the stress-tensor
corresponding to the two corrective Lagrangians (\ref{eq:lcor},
\ref{eq:lu}), and the equations of motion of the scalar field and the
unit vector field. Having obtained these, we discuss their properties 
in the FRW background.

This question has been addressed in Ref. \cite{Lemoine-Lubo:2001}. Our
treatment differs with the one followed in that paper by the fact that
we do not make the decomposition in time and space-like
components. Our condensed formulas will prove very useful when
computing the perturbations.

In order to do that, it proves easier to go back to the action. Let us
first specialize to the contribution of the non-linear part of the
Lagrangian (\ref{eq:lcor}):
\be
\label{eq:app-lcor}
S_{_{\rm cor}} = - b_{_{11}} \int d^4 x \sqrt{-g} ( {\cal D}^2
\varphi)^2  \quad .
\ee
Using the definition given in Eq.(\ref{eq:DDvp}), the variation of
${\cal D}^2 \varphi$ can be written as
\begin{eqnarray}
\delta ( {\cal D}^2 \varphi) &=& \wt{A}_{\mu \nu} \delta g^{\mu \nu} +
\wt{B}^\sigma_{\mu \nu}  \partial_\sigma \delta g^{\mu \nu} +
\wt{C}^\mu \delta u_\mu + \wt{D}^{\nu \mu} \partial_\nu \delta u_\mu \, .
\nonumber\\
%%%%
\label{eq:del-Dvp}
& & + \wt{E}^\mu \partial_\mu \delta \varphi + 
\wt{H}^{\mu \nu} \partial_\mu \partial_\nu  \delta \varphi  \, .
\end{eqnarray}
The quantities $\wt{A} \cdots \wt{H}$ can be written explicitly.  For
example,
\be
\wt{A}_{\mu \nu} = \frac{\partial}{\partial g^{\mu \nu}} {\cal D}^2
\varphi \, , \quad 
\wt{H}^{\mu \nu} = \frac{\partial}{\partial 
(\partial_\mu \partial_\nu \varphi)} {\cal D}^2 \varphi  \, .
\ee
These partial derivatives have to be taken keeping in mind the choice
of variables made in this work. In order to be consistent, we choose
the following set of independent variables
\br
& & \varphi, \pa_{\b}\varphi, \pa_{\a}\pa_{\b}\varphi, 
\pa_{\s}\pa_{\a}\pa_{\b}\varphi, 
g^{\mu\nu}, \pa_{\rho} g^{\mu\nu}, \pa_{\tau} \pa_{\rho} g^{\mu\nu},
\nn \\ 
& &  \qquad \qquad u_{\sigma}, \pa_{\b} u_{\sigma}, \pa_{\a}\pa_{\b} 
u_{\sigma}, \lambda  \, .
\label{eq:cho-var}
\er

Using the relations
%From relations such as
%
\be
\partial_\ve ( g_{\vk\vr} g^{\vr\vs}) = 0 \quad ; \quad 
\frac{\partial}{\partial g_{\alpha \beta}} (g_{\vk\vr} g^{\vr\vs} ) = 0 
\, ,
\label{eq:iden} 
\ee
%one finds the following formulas which are needed to achieve the computations 
we get,
\begin{eqnarray}
  \frac{\partial g^{\alpha \beta} }{\partial g^{\rho \sigma}} &=&
   \delta_\rho^\alpha \delta_\sigma^\beta  \, , \quad 
\frac{\partial g_{\alpha \beta} }{\partial g^{\rho \sigma}} =
  - g_{\rho  \alpha} g_{\sigma  \beta}   \, , \nonumber\\
\partial_\epsilon g_{\alpha \beta} &=& - g_{\alpha \rho} g_{\beta \sigma} 
\partial_\epsilon g^{\rho \sigma} \quad .
\end{eqnarray}
The expressions giving the quantities $\wt{A}, \cdots, \wt{H}$ are
very long and time consuming to obtain and will not be displayed
here. One of the most important aspects of the way we will present our
results is that for our purposes we only need to know their values on
the FRW background. This is the subject of the next appendix.

%Let us come back to the action.
Substituting Eq. (\ref{eq:del-Dvp}) in (\ref{eq:app-lcor}) and
integrating by parts the resultant expression, we obtain
{\small
\begin{eqnarray}
%\label{}
\delta S_{_{cor}}&=&  - b_{11} \int d^4 x \left[ \sqrt{-g} \left(
\frac{1}{2} g_{\mu \nu} ({\cal D}^2 \varphi)^2 - 2 \wt{A}_{\mu \nu}
{\cal D}^2 \varphi \right) \right. \nonumber\\ 
%%%%%
& & \qquad \qquad \qquad  \qquad \qquad + \left. \partial_\sigma
\left( \sqrt{-g} {\cal D}^2 \varphi \wt{B}^\sigma_{\mu \nu} \right) \right]
\delta g^{\mu \nu} \nonumber\\
%%%%%%
& & - 2 b_{11} \int d^4 x \left[ \sqrt{-g} {\cal D}^2 \varphi \wt{C}^\mu -  
  \partial_\nu 
  \left( \sqrt{-g} {\cal D}^2 \varphi \wt{D}^{\nu \mu} \right)
  \right] \delta u_\mu  \nonumber\\
%%%%%%
%\label{}
& & - 2 b_{11} \int d^4 x \left[ - \partial_\mu \left( \sqrt{-g} 
{\cal D}^2 \varphi \wt{E}^\mu \right) \right. \\ 
%%%%%
& & \qquad \qquad \qquad  + \left.
\partial_\mu \partial_\nu \left( \sqrt{-g} {\cal D}^2 \varphi 
 \wt{H}^{\mu \nu} \right) \right] \delta \varphi \nonumber \quad .
\end{eqnarray}
}
%
%From here  can infer the contribution to the stress-energy tensor as well as
%the contributions to the field equations of the inflaton and the vector field 
%$u$:
\indent From the above expression, it is easy to infer the contribution of the
corrective Lagrangian to the stress-energy tensor as well as to the
field equations of the inflaton and the vector field $u_{\mu}$, i. e.,
{\small
\begin{eqnarray}
\label{eq:tmu-app}
T^{({\rm cor})}_{\mu \nu}&=& - b_{_{11}} g_{\mu \nu} ({\cal D}^2 \varphi)^2 
- 4  b_{_{11}} \wt{B}^\rho_{\mu \nu} \partial_\rho ({\cal D}^2 \varphi) 
\nn \\ 
%%%%%%%
&+&  b_{_{11}} \left( 4 \wt{A}_{\mu \nu} - 2 \frac{\partial_\rho g}{g} 
\wt{B}^\rho_{\mu \nu} 
- 4 \partial_\rho \wt{B}^\rho_{\mu \nu} \right) {\cal D}^2 \varphi \\
%%%%%%%
& \equiv &  b_{_{11}} g_{_{\mu \nu}} ({\DD}^2 \varphi)^2  
+  b_{_{11}} E_{_{\mu\nu}} \DD^2\varphi 
+ 4 b_{_{11}} C_{_{\mu \nu}}^{\rho \vk}  \partial_{\vk}\varphi  \, 
\partial_\rho[{\DD}^2 \varphi] \, , \nn \\
%%%%%%%%%%%%%%
\label{eq:eq1vp}
eq_{1,\varphi}&=&  \frac{2 b_{_{11}}}{\sqrt{-g}} 
\left( \left[ \partial_\mu \left( \sqrt{-g} \wt{E}^\mu \right)  
- \partial_\mu \partial_\nu \left( \sqrt{-g} \wt{H}^{\mu \nu} \right)
 \right] {\cal D}^2 \varphi \right. \nonumber\\
%%%%%
&+& \left. \left[ \sqrt{-g} \wt{E}^\mu - 
\partial_\nu \left( \sqrt{-g} \wt{H}^{\mu \nu} \right)
- \partial_\nu \left( \sqrt{-g} \wt{H}^{\nu \mu} \right) \right] 
\partial_\mu {\cal D}^2 \varphi \right. \nonumber\\
%%%%%
& & \left. - \sqrt{-g} \wt{H}^{\mu \nu}  \partial_\mu \partial_\nu 
{\cal D}^2 \varphi \right)  \, , \\
%%%%%%%%%%%%%%%%
\label{eq:eq2vp}
eq_{2,\varphi} &=& - 2 b_{11} \left[  \left( \wt{C}^\mu - \frac{1}{2} \frac{\partial_\nu g}{g} \wt{D}^{\nu \mu}
 - \partial_\nu \wt{D}^{\nu \mu} \right) {\cal D}^2 \varphi \right. 
 \nonumber\\
& & \left. 
- \wt{D}^{\nu \mu} \partial_\nu {\cal D}^2 \varphi \right]
 \, ,
\end{eqnarray}
}
where (\ref{eq:eq1vp}) gives the RHS of Eq. (\ref{eq:DE-vp}) and 
(\ref{eq:eq2vp}) gives the RHS of Eq. (\ref{eq:DE-u}).

%In the same way, the action containing the vector field leads to
Similarly, the variation of the action for the unit vector field
(\ref{eq:lu}) leads to
\begin{eqnarray}
%\label{}
\delta S_u &= & \int d^4 x \sqrt{-g} T_{\mu \nu} \delta g^{\mu \nu} 
- 2 \lambda \int d^4 x \sqrt{-g} g^{\mu \nu} u_\nu \, \delta u_\mu \nonumber\\
%%%%%%%%
&+& \int d^4 x  \sqrt{-g} (- g^{\alpha \beta} u_\alpha u_\beta + 1) \,
\delta \lambda \nonumber\\
%%%%%%%
&-& d_1 \int d^4 x \,
\sqrt{-g} g^{\alpha \rho} g^{\beta \rho} \delta( F_{\alpha \beta} 
F_{\rho \sigma} )  \quad .
\end{eqnarray}
\indent From the above expression, we obtain the contribution to the
stress-energy tensor and the field equation of the vector field, i. e.,
\begin{eqnarray}
T^{(u)}_{\mu \nu} &=&  \left( \frac{1}{2} 
\lambda ( g^{\alpha \beta} u_\alpha u_\beta -1) + \frac{1}{2}
d_1 F_{\alpha \beta} F^{\alpha \beta} \right) g_{\mu \nu}  \nonumber\\
%%%%
&-&  \lambda u_\mu u_\nu - 2 d_1 g^{\alpha \rho} F_{\alpha \mu} 
F_{\rho \nu}  \, , \\ 
%%%%%%%%%%%%%
\label{eq:eq1u}
eq_{2,u} &=& -2 \lambda g^{\mu \nu} u_\nu - 2 d_1  \frac{1}{\sqrt{-g}}
\pa_{\nu} \l[ \sqrt{-g} F^{\mu\nu}\r] \, ,
%+ 2 d_1 \frac{1}{\sqrt{-g}}
%\partial_\nu \left[ \sqrt{-g} ( g^{\rho \nu} g^{\sigma \mu} 
%\right.  \nonumber\\
%& & \left.  -  g^{\rho \mu} g^{\sigma \nu} ) 
%    F_{\rho \sigma} \right]
 \, .
\end{eqnarray}
where (\ref{eq:eq1u}) gives the LHS of Eq. (\ref{eq:DE-u}).

It is easy to verify that the energy momentum tensor corresponding to
the corrective Lagrangian and the vector field vanish while the
equation of motion of the vector field and the inflaton are 
satisfied when the following relation holds
\begin{equation}
\label{eq:bg-u}
\ov{u}_{\mu} = a(\eta)  \l(1, 0, 0, 0\r)~~,~~
\ov{\lambda} =  0~~,~~ 
\pa_{\mu}\pa^{\mu}{\ov{\vp}}= \frac{{\ov{\vp}^{'}}^2}{a^2(\eta)} 
\, .
\end{equation}
The intermediary relations needed are
\begin{eqnarray}
\label{eq:bg-DD}
 \ov{F}_{\mu \nu} = 0~,~
\ov{\perp}_{\a\b} = a^2 \l(\eta_{\a\b} - \d^0_{_\a} \d^0_{_\b}\r) \, , \,
\DD^2\l(\ov{\vp}\r) = 0  \, .~~~ 
\end{eqnarray}
The evolution of the modified scalar field in the FRW background is
same as that of the inflaton.  Hence, the trans-Planckian effects, in
this model, only affect the perturbations.  As we will see in the next
appendix, the above equations play a significant role in discarding
many terms from the perturbed stress-tensor.

\section{The Perturbed FRW space time.}

In this appendix, we obtain the linear order perturbed stress-tensor
corresponding to the two corrective Lagrangians (\ref{eq:lcor},
\ref{eq:lu}), and the equations of motion of the perturbed scalar field 
and the unit vector field.

%Let us begin by the Einstein equations. 
We will use more than once the Leibniz rule:
\begin{eqnarray}
  \delta( X Y) &=& \ov{X} \delta Y + \ov{Y} \delta X 
  \equiv X_o \delta Y + Y_o \delta X  \, .
\end{eqnarray}
and the vanishing of the three dimensional Laplacian of the scalar
field in the FRW background.

For example, the variation of the component of the stress-energy
tensor coming from the non-linear part of the Lagrangian reads
\begin{eqnarray}
\delta T^{corr}_{\mu \nu} &=& 
\left[ 4 \wt{A}_{\mu \nu} - 2 \frac{\partial_\rho g}{g} 
\wt{B}^\rho_{\mu \nu}
- 4 \partial_\rho \wt{B}^\rho_{\mu \nu} \right]_o   
\delta {\cal D}^2 \varphi \nonumber\\
%%%%%
&-& 4 \left( \wt{B}^\rho_{\mu \nu} \right)_o \partial_\rho 
( \delta {\cal D}^2 \varphi)  \quad .
\end{eqnarray}
\indent From the above expression, we see that it is enough to know the
quantities $\wt{A}, \cdots, \wt{H}$ on the background. Going back to
Eq. (\ref{eq:del-Dvp}), we have to compute the variation of the vector
field, the Christoffel symbol and other quantities. Let us begin by
the connection coefficient:
\begin{eqnarray}
\Gamma^\alpha_{\rho \sigma} = \frac{1}{2} g^{\alpha \tau}
( - \partial_\tau g_{\rho \sigma} + \partial_\rho g_{\sigma \tau} +
\partial_\sigma g_{\tau \rho})   \quad .
\end{eqnarray} 
Due to our choice of independent variables, we need to express the
variations of the covariant components of the metric in terms of the
contravariant ones.  This is achieved simply:
\be
\delta (g_{\alpha \beta} g^{\beta \gamma}) = 0 \Longrightarrow
\delta g_{\alpha \beta} = - g_{\alpha \gamma} g_{\beta \sigma}
  \delta g^{\gamma \sigma} \quad .
\ee
One then obtains
\begin{eqnarray} 
\delta \Gamma^\alpha_{\rho \sigma}&=&  a \, a' \, 
\l( - \delta^\alpha_\mu \delta^0_\nu \eta_{\rho \sigma} + 
\delta^\alpha_\mu  \eta_{\sigma \nu} \delta^0_\rho
- \delta^\alpha_\mu \eta_{\rho \nu} \delta^0_\sigma \r.  \nonumber\\
%%%%
& &  \qquad +\l. 2 \delta^\alpha_0 \eta_{\rho \mu} \eta_{\sigma \nu} -
2  \delta^\alpha_\nu \delta^0_\rho \eta_{\mu \sigma} \r)  
\delta g^{\mu \nu} \\
%%%%%%%
& +& \frac{a^2}{2} 
\l(\eta_{\rho \mu} \eta_{\sigma \nu} \eta^{\alpha \epsilon} -
\eta_{\sigma \mu} \delta^\alpha_\nu \delta^\epsilon_\rho -
\eta_{\rho \nu} \delta^\alpha_\mu \delta^\epsilon_\sigma  \r) 
\partial_\epsilon \delta g^{\mu \nu} \,. \nn
\end{eqnarray} 
Similarly, we get
\begin{eqnarray}
\delta\perp^{\rho \sigma} &=& ( - \delta^\rho_\mu \delta^\sigma_\nu + 
\delta^\sigma_0 \delta^\rho_\mu \delta^0_\nu +
+ \delta^\rho_0 \delta^\sigma_\mu \delta^0_\nu ) \delta g^{\mu \nu} \nonumber\\
%%%%%
&+& ( \delta^\rho_0 \eta^{\sigma \mu} + \delta^\sigma_0 \eta^{\rho \mu} ) 
\delta u_\mu \, .
\end{eqnarray}
After a long but straightforward calculation, one obtains the result
given in Eqs.~(\ref{eq:Def-Emn},\ref{eq:Def-C} ) with the following
expressions for the background:
\begin{eqnarray}
(\wt{A}_{\mu \nu})_o = 
\left( \bar{\varphi}^{''} + 3 \frac{a^{'}}{a} \bar{\varphi}{'} 
\right) \delta^0_\mu \delta^0_\nu &, &
(\wt{B}^\epsilon_{\mu \nu})_o = \frac{1}{2}  \bar{\varphi}^{'}
\delta^0_\mu \delta^0_\nu \delta_0^\epsilon \, , \nonumber\\
%%%%%%
(\wt{C}^{\mu})_o = \frac{1}{a^3} \left(2 \bar{\varphi}^{''} + 
3 \frac{a^{'}}{a} \bar{\varphi}^{'} \right) \delta_0^\mu &,&
(\wt{D}^{\sigma \rho})_o = 
\frac{1}{a^3} \bar{\varphi}^{'} \eta^{\rho \sigma} \, ,  \nonumber\\
%%%%%
\label{eq:bABCE1}
(\wt{H})_o = \frac{1}{a^3} 
( - \eta^{\mu \nu} + \delta^\mu_0 \delta^\nu_0)  & ,&   
(\wt{E})_o = 0 \, .
\end{eqnarray}
At this point it is worth noticing that $0-0$ is the only non-zero
component of the stress-tensor. Since the trans-Planckian corrections
do not change the pressure perturbations, the perturbation equation
for the tensor modes do not change. Hence, the tensor perturbation
equations is given by Eq. (\ref{eq:TPer0}). Recently, Lim
\cite{Lim:2004} had show that general Lorentz violating models (with
out taking into account higher derivatives of the scalar field) can
modify the pressure perturbations and hence the tensor perturbation
equations. However, in our specific Lorentz violating model, this is
not the case.

Let us now specialize to the scalar perturbations. The covariant
components read
\beq
\label{eq:scalar-per}
\d g_{\mu\nu} = a^{2}(\eta) 
\left(\begin{array}{ccc}
2 \phi & & - \pa_i B \\
- \pa_i B & & 2 \psi \d^{ij} - 2 \pa_i \pa_j E
\end{array}\right) \, .
\eeq
Using the identities given in Eq.~(\ref{eq:iden}), we get the
contravariant components
{\small
\br
\!\!\!
\d g^{00} = \frac{- 2 \phi}{a^2(\eta)},~
\d g^{0i} = \frac{\pa_i B}{a^2(\eta)},~ 
%\nn \\ & & 
\d g^{ij} = \frac{\l(2 \pa_i\pa_j E - 2 \psi \d_{ij}\r)}{a^2(\eta)} \, .
\er
}
Using the fact that $u_{\mu}$ is a unit vector, we have
\be
  \delta (g^{\alpha \beta} u_\alpha u_\beta) = 0  \Longrightarrow 
  \delta u_0 =  a \phi \quad .
\ee
Using the form of the scalar perturbations Eq.~(\ref{eq:scalar-per})
and the results obtained in Eq.(\ref{eq:bABCE1}), we obtain the
following simple formula
\be
\label{eq:d-DD} 
\delta ( {\cal D}^2 \varphi) = \frac{1}{a^2} \left( - \frac{1}{a}
 \bar{\varphi}^{'} \delta^{ij} \partial_i \delta u_j + \nabla^2 \delta
 \varphi \right) \, .
\ee
Let us now turn our attention to the equation of motion obeyed by the
perturbations. Concerning the inflaton field, the first two
contributions to the formula given in Eq.~(\ref{eq:eq1vp}) vanish and
the result is
\be
\delta eq_{1,\varphi} =  \frac{2 b_{11}}{a^4}
\left[ - \nabla^4 (\delta \varphi) + \frac{1}{a} \bar{\varphi}^{'} 
 \nabla^2 ( \delta^{ij} \partial_i \delta u_j )   \right] \, .
\ee 
In the same way, we get
\begin{eqnarray}
& & b_{11} \frac{1}{a^6} \delta^\mu_0 \left( 10 \frac{a^{'}}{a}  
( \bar{\varphi}^{'})^2  \delta^{ij} \partial_i \delta u_j 
  - ( 2 \bar{\varphi}^{''} + 8 \frac{a^{'}}{a}  \bar{\varphi}^{'}) a  
  \nabla^2 \delta \varphi \right) \nonumber\\
%%%%%
&+& b_{11} \frac{1}{a^6} \eta^{\mu \nu} \bar{\varphi}^{'}
( 2 a \nabla^2 \partial_\nu  \delta \varphi - 2 \bar{\varphi}^{'} \partial_\nu 
\delta^{ij} \partial_i \delta u_j)
- 2 \frac{1}{a} \delta^\mu_0 \delta \lambda   \nonumber\\
%%%%%
&+& 4 d_1 \frac{1}{a^4} ( \eta^{\rho \nu} \eta^{\sigma \mu} - 
\eta^{\rho \mu} \eta^{\sigma \nu}) \partial_\nu \partial_\rho 
\delta u_\sigma  = 0  \, .
\end{eqnarray}
From the temporal index ($\mu=0$), we obtain the variation of the
Lagrange multiplier, i. e.,
\begin{eqnarray}
\delta \lambda &=& 2 d_1 \frac{1}{a^3} \delta^{ij} \partial_j (\delta u_i)^{'} 
+ \frac{b_{11}}{a^5} \l[ a \bar{\varphi}' \nabla^2 \delta \varphi' - 
\ov{\varphi}'^2  \delta^{ij} \partial_i \delta u_j' \r. \nn \\
%%%
& & + \left. 5 {\cal{H}}
\ov{\varphi}'^2 \delta^{ij} \partial_j \delta u_i - 
a\, \l(\ov{\varphi}'' + 4 {\cal{H}} \bar{\varphi}^{'} \r) \nabla^2
\delta \varphi \right] \, .
\label{eq:p-lam} 
\end{eqnarray}
From the spatial indices $\mu=k$, we obtain
\begin{eqnarray}
& & b_{11} \left(-2 a \bar{\varphi}^{'} \nabla^2 \partial_k 
\delta\varphi + 2 (\bar{\varphi}^{'})^2  \delta^{ij} 
\partial_k \partial_i \delta u_j \right) \\
%%%%%%
&+& 4 d_1 a^2 \l[ - \delta u_k'' + \nabla^2 \delta u_k + 
\partial_k (a \phi)^{'}
- \partial_k \partial_i \delta u_i \r] = 0  \, . \nn
\end{eqnarray}
Having these results, especially the variation of the Lagrange
multiplier, we obtain the expression for the non-vanishing component
of the stress-tensor:
\begin{eqnarray}
\delta T_{00}^{\rm (cor)} &=& \frac{2 b_{11}}{a^2} \left(5
\frac{{\cal H}}{a} \ov{\varphi}'^2 \partial_i \delta u_i -
\frac{1}{a} \ov{\varphi}'^2 \partial_i \delta u_i'
\right. \\
%%%%
& & \left. -~ (\ov{\varphi}'' + 4 {\cal H} \ov{\varphi}') 
\nabla^2 \delta \varphi + \ov{\varphi}'\nabla^2\delta \varphi' 
\right) \, , \nonumber \\
\delta T^{u}_{00}& =& - 2 a^2 \delta \lambda \, .
\end{eqnarray}
We now wish to express all the quantities in terms of gauge invariant
quantities. For the vector field, we have
\be
\delta u_i = \delta u_i^{(gi)} - u_0 \, \partial_i (B - E^{'})  \, .
\ee
The equation of the vector field now becomes
{\small
\begin{eqnarray}
&& \!\!\!\!\!\!\! - b_{_{11}}  
\left(a \, \ov{\varphi}' \nabla^2{\partial_k \delta \varphi^{(gi)}}
- {\ov{\varphi}'}^2  \delta^{ij} \partial_k \partial_i 
\delta u^{(gi)}_j \right) \\
%%%%%
& &  \!\!\!\!\!\!\!
+ 2 d_1 a^2 \left(-{\delta u^{(gi)}_k}'' + \nabla^2 \delta u^{(gi)}_k 
+ \partial_k (a \Phi)' 
+ \partial_k \partial_i \delta u^{(gi)}_i \right) = 0  \, , \nn
\end{eqnarray}
}

\noindent while for the inflaton one obtains
{\small
\begin{eqnarray}
& & \!\!\!\!\!\!\!\!\!
2 V_{,\vp} a^2 \Phi - 4 \ov{\varphi}^{'} \Phi^{'} 
+ a^2 V_{,\vp\vp} \delta \varphi^{(gi)} 
+ 2 {\cal H} {\delta \varphi^{(gi)}}' +
{\delta \varphi^{(gi)}}''  \\
%%%%%
& & \!\!\!\!\!\!\!\!\!
-  \nabla^2 \delta \varphi^{(gi)} %\nonumber\\
% &+& 
+ \frac{2 b_{11}}{a^2} \left[ - \nabla^4 \delta \varphi^{(gi)} +
 \frac{1}{a} \bar{\varphi}^{'} \nabla^2 \delta^{ij} \partial_i 
 \delta u_j^{(gi)}  
 \right] = 0 \, .~~~~~~ \nn
\end{eqnarray}
}
Similarly, we can obtain expressions for the stress-energy tensor
(\ref{eq:per-fin}) interms of the gauge-invariant variables.

Until this point we have not assumed any specific form of the spatial
part of the perturbed unit-vector field. In general, the
time-dependent spatial components of the perturbed $u$ field can be
expressed as a sum of irrotational and solenoidal parts, i. e.
\beq
\d u_i \equiv (\nabla \xi)_i + (\nabla \times \varpi)_i \, .
\eeq
The irrotational part of the perturbed $u$ field will contribute to
the scalar perturbations while the solenoidal part of the perturbed
$u$ field contributes to the vector perturbations. Since we 
are interested in the scalar perturbations, for the rest 
of the calculations, we ignore the solenoidal part. Thus, we get
%We are dealing with scalar perturbations which means the
%only terms appearing in the metric are scalars $(\phi,\psi)$ or
%derivatives of scalars ($B,E$). Willing the spatial part of the field
%responsible for trans-Planckian physics to be a derivative
%\footnote{ .}, we get
%
\be
\label{eq:def-xi}
\delta u^{(gi)}_i =  \partial_i \xi^{(gi)}  \, .
\ee
Substituting this in the Einstein and the field equations, we obtain
Eqs.~(\ref{eq:per-fin}).

%%%%%%%%%%%% NEW APPENDIX %%%%%%%%%%%%%%%%%%%%%%%%%

\section{The fourth order equation}

In this appendix, we provide key steps in obtaining the equation of
motion of Bardeen potential ($\Phi$), in a general FRW background, by
combining Eqs. (\ref{eq:MDR-Phi},\ref{eq:per-vpTP},\ref{eq:per-uTP}).

From the constraint equation we deduce that
\be
 \delta \varphi_k = \frac{2 }{3} M_{pl}^2 \left(
 \frac{1}{\bar{\varphi}^{'}} \Phi^{'}_k +  
 \frac{{\cal H}}{\bar{\varphi}^{'}} \Phi_k  \right) \, .
\ee
Substituting this expression and its derivatives into the equation of
motion of $\delta \varphi_k$, one obtains $\xi_k$, i. e.,
\be
\xi_k = \frac{M_{pl}}{k^2} 
\left( Q_3 \Phi_k^{'''} + Q_2 \Phi_k^{''} + Q_1 \Phi_k^{'} + Q_0 \Phi_k^{} 
\right) \, ,
\ee
where
\begin{eqnarray}
Q_3 &=& \frac{M_{pl}^2}{3 b_{11}}  \frac{a^3}{\bar{\varphi}^{'2}}  
\frac{1}{k^2}\, , \, 
Q_2 = \frac{M_{pl}^2}{3 b_{11}} \frac{a^3}{\bar{\varphi}^{'}}
\left[ 2 \left( \frac{1}{\bar{\varphi}^{'}} \right)^{'} + 
\frac{\cal H}{\bar{\varphi}^{'}}   \right]  \frac{1}{k^2} \, ,   \nonumber\\
%%%%
Q_1 &=& \left[ \frac{M_{pl}^2}{3 b_{11}} \frac{a^3}{\bar{\varphi}^{'}}
  \left(  \left( \frac{1}{\bar{\varphi}^{'}} \right)^{''} + 
  2 \left( \frac{\cal H}{\bar{\varphi}^{'}} \right)^{'} - 
  2 \frac{1}{b_{11}} a^3 \right) \right] \frac{1}{k^2}  \, , \nonumber\\
%%%
&+& \frac{M_{pl}^2}{3 b_{11}}  \frac{a^3}{\bar{\varphi}^{'2}}
  + \frac{2}{3} M_{pl}^2  \frac{a}{\bar{\varphi}^{'2}} k^2 \, , \nonumber\\
%%%%%%  
Q_0 &=& \left[ \frac{M_{pl}^2}{3 b_{11}} \frac{a^3}{\bar{\varphi}^{'}}
\left( \frac{{\cal H}}{\bar{\varphi}^{'}} \right)^{''} +
\frac{1}{b_{11}} \frac{a^5}{\bar{\varphi}^{'}} \bar{V}_{,\varphi} \right] 
\frac{1}{k^2} + \frac{M_{pl}^2}{3 b_{11}} 
\frac{a^3}{\bar{\varphi}^{'2}} {\cal{H}} \nonumber\\
%%%
&+& \frac{2}{3} M_{pl}^2  \frac{a}{\bar{\varphi}^{'2}} {\cal{H}} \, k^2  \, .
\end{eqnarray} 
Substituting the above expressions in the field equation of the
Bardeen potential, we get,
\begin{eqnarray} 
& & \Phi^{''''}_k + X_3 \Phi^{'''}_k + ( X_2 + Y_2 k^2 + Z_2 k^4 ) 
\Phi^{''}_k  \\
&+& ( X_1 + Y_1 k^2 + Z_1 k^4 ) \Phi^{'}_k  %\nonumber\\
+ ( X_0 + Y_0 k^2 + Z_0 k^4 ) \Phi_k = 0 \, , \nn 
\end{eqnarray} 
where
\begin{eqnarray} 
X_3 &=& \frac{\ov{\varphi}^{'2}}{a^3}   
 \left( \frac{a^3}{\ov{\varphi}^{'2}} \right)^{'} +
 \ov{\varphi}^{'}  \left[  2 \left( \frac{1}{\ov{\varphi}^{'}} \right)^{'} 
   + \frac{{{\cal H}}}{\ov{\varphi}^{'}} \right] \, , \nonumber \\
%%%%%%%
 X_2 &=& \frac{\ov{\varphi}^{'2}}{a^3}
 \left[ \frac{a^3}{\ov{\varphi}^{'}} \left(  2 \left( \frac{1}{\ov{\varphi}^{'}} \right)^{'} 
   + \frac{{{\cal H}}}{\ov{\varphi}^{'}} \right) \right]^{'} \nonumber \\
%%%%
&+& \ov{\varphi}^{'}  \left[ \left( \frac{1}{\ov{\varphi}^{'}} \right)^{''} 
   +2 \left( \frac{{{\cal H}}}{\ov{\varphi}^{'}} \right)^{'} \right]
   - 6 \frac{1}{M_{pl}^2} \ov{\varphi}^{'2} \, , \nonumber %\\
%%%%%%%
\end{eqnarray}
%%%
\begin{eqnarray}
Y_2 &=& \frac{3}{2} \frac{b_{11}}{d_1} \frac{\ov{\varphi}^{'2}}{a^3} +1 
\, , \quad Z_2 = 2 b_{11} \frac{1}{a^2} \, , \nonumber \\
%%%%%%
X_1 &=& \frac{\ov{\varphi}^{'2}}{a^3} \left[
\frac{a^3}{\ov{\varphi}'} 
\left(\left( \frac{1}{\ov{\varphi}^{'}} \right)^{''} 
+ 2 \left(\frac{{{\cal H}}}{\ov{\varphi}^{'}} \right)^{'} \right) 
- 6 \frac{1}{M_{pl}^2} a^3 \right]^{'} \, , \nonumber \\
%%%%
Y_1 &=&  \frac{\ov{\varphi}^{'2}}{a^3}   
 \left( \frac{a^3}{\ov{\varphi}^{'2}} \right)^{'} + {\cal H} 
 + 3 \frac{b_{11}}{d_1} \frac{\ov{\varphi}^{'2}}{a^3} \left( {\cal H} - 
 \frac{\ov{\varphi}^{''}}{\ov{\varphi}^{'}}   \right) \, , \nonumber\\
%%%%
Z_1 &=& 2 b_{11} \frac{\ov{\varphi}^{'2}}{a^3}   
 \left[ \left( \frac{a}{\ov{\varphi}^{'2}} \right)^{'} +
  \frac{a}{\ov{\varphi}^{'2}}  {\cal H}  
 \right]  \, , \\
%%%%%
X_0 &=& \frac{\ov{\varphi}^{'2}}{a^3}   
 \left[  \frac{a^3}{\ov{\varphi}^{'}} 
 \left( \frac{{\cal H}}{\ov{\varphi}^{'}} \right)^{''}
 + 3 \frac{1}{M_{pl}^2} \frac{a^5}{\ov{\varphi}^{'}} \ov{V}_{,\varphi}  
 \right]^{'}  \, , \nonumber\\
%%%%%%
Y_0 &=& \frac{\ov{\varphi}^{'2}}{a^3}  
\left( \frac{a^3}{\ov{\varphi}^{'2}} {\cal H} \right)^{'} +
3 \frac{b_{11}}{d_{1}}  \frac{\ov{\varphi}^{'2}}{a^3}
\left( {\cal H}^{'} - {\cal H}  
\frac{\ov{\varphi}^{''}}{\ov{\varphi}^{'}}  \right)  \, , \nonumber\\
%%%%%
Z_0 &=& 2 b_{11} \frac{{\ov{\varphi}'}^{2}}{a^3}   
\left( \frac{a'}{{\ov{\varphi}'}^{2}}\right)^{'} 
%\nonumber\\
%&+& 
+ \frac{3}{2} b_{11}  \left( \frac{1}{d_1} - 2 \frac{1}{M_{pl}^2}  \right)
 \frac{\ov{\varphi}^{'2}}{a^3} \, . \nn
\end{eqnarray}

%%%%%%%%%%%%%%%%%%%%%%%%%%%%%%%%%%%%%%%%%%%%%%%%%%%%%%%%%%%%%%%%%%%%%%%%%%
%%%%%%%%%%%%%%%%%%%%%%%% BIBLIOGRAPHY %%%%%%%%%%%%%%%%%%%%%%%%%%%%%%%%%%%%
%%%%%%%%%%%%%%%%%%%%%%%%%%%%%%%%%%%%%%%%%%%%%%%%%%%%%%%%%%%%%%%%%%%%%%%%%%

\end{document}